\documentclass[lettersize,hidelinks,journal]{IEEEtran}

\usepackage{cite}
\usepackage{amsmath,amssymb,amsfonts}
\usepackage{graphicx}
\usepackage{textcomp}
\usepackage[dvipsnames]{xcolor}
\usepackage[acronym,shortcuts]{glossaries}
\usepackage{bm}
\usepackage{caption}
\usepackage{subcaption}
\usepackage{relsize}
\usepackage{soul}
\usepackage{algorithm, algpseudocode}
\usepackage{algcompatible}
\usepackage{outlines}
\usepackage{orcidlink}
\usepackage{lipsum,comment}
\usepackage{mathtools}
\usepackage{tabularx,flushend}
\usepackage{amsthm}
\newtheorem{remark}{Remark}

\theoremstyle{definition}
\newtheorem{lemma}{Lemma}

\def\BibTeX{{\rm B\kern-.05em{\sc i\kern-.025em b}\kern-.08em
T\kern-.1667em\lower.7ex\hbox{E}\kern-.125emX}}

\newcommand{\trans}[0]{^{\mathsf{T}}}

\newcommand{\herm}[0]{^{\mathsf{H}}}

\newacronym{RPE}{RPE}{radar parameter estimation}
\newacronym{OTFS}{OTFS}{orthogonal time frequency space}
\newacronym{AFDM}{AFDM}{affine frequency division multiplexing}
\newacronym{MIMO}{MIMO}{multiple-input multiple-output}
\newacronym{SISO}{SISO}{single-input single-output}
\newacronym{ISAC}{ISAC}{integrated sensing and communications}
\newacronym{3D}{3D}{three-dimensional}
\newacronym{2D}{2D}{two-dimensional}
\newacronym{1D}{1D}{one-dimensional}
\newacronym{RX}{RX}{receiver}
\newacronym{TX}{TX}{transmitter}
\newacronym{BF}{BF}{beamforming}
\newacronym{mmWave}{mmWave}{millimeter-wave}
\newacronym{SotA}{SotA}{state-of-the-art}
\newacronym{ULA}{ULA}{uniform linear array}
\newacronym{QAM}{QAM}{quadrature amplitude modulation}
\newacronym{ISFFT}{ISFFT}{inverse symplectic finite Fourier transform}
\newacronym{SFFT}{SFFT}{symplectic finite Fourier transform}
\newacronym{AWGN}{AWGN}{additive white Gaussian noise}
\newacronym{OFDM}{OFDM}{orthogonal frequency division multiplexing}
\newacronym{OCDM}{OCDM}{orthogonal chirp division multiplexing}
\newacronym{BS}{BS}{base station}
\newacronym{UE}{UE}{user equipment}
\newacronym{DFT}{DFT}{discrete Fourier transform}
\newacronym{IDFT}{IDFT}{inverse discrete Fourier transform}
\newacronym{TD}{TD}{time-domain}
\newacronym{wlg}{wlg}{without loss of generality}
\newacronym{CP}{CP}{cyclic prefix}
\newacronym{DAFT}{DAFT}{discrete affine Fourier transform}
\newacronym{IDAFT}{IDAFT}{inverse discrete affine Fourier transform}
\newacronym{CPP}{CPP}{\textit{chirp-periodic} prefix}
\newacronym{IDZT}{IDZT}{inverse discrete Zak transform}
\newacronym{DZT}{DZT}{discrete Zak transform}
\newacronym{ICI}{ICI}{inter-carrier interference}
\newacronym{BER}{BER}{bit error rate}
\newacronym{DoF}{DoF}{degrees-of-freedom}
\newacronym{FD}{FD}{full-duplex}
\newacronym{SIMO}{SIMO}{single-input multiple-output}
\newacronym{MISO}{MISO}{multiple-input single-output}
\newacronym{AoD}{AoD}{angle-of-departure}
\newacronym{AoA}{AoA}{angle-of-arrival}
\newacronym{RF}{RF}{radio frequency}
\newacronym{SIM}{SIM}{stacked intelligent metasurfaces}
\newacronym{FIM}{FIM}{flexible intelligent metasurface}
\newacronym{FPGA}{FPGA}{field programmable gate array}
\newacronym{UPA}{UPA}{uniform planar array}
\newacronym{CC}{CC}{communication-centric}
\newacronym{I/O}{I/O}{input-output}
\newacronym{iid}{i.i.d.}{independent and identically distributed}
\newacronym{IoT}{IoT}{internet of things}
\newacronym{V2X}{V2X}{vehicle-to-everything}
\newacronym{NTN}{NTN}{non-terrestrial network}
\newacronym{LEO}{LEO}{low-earth orbit}
\newacronym{THz}{THz}{terahertz}
\newacronym{EM}{EM}{electromagnetic}
\newacronym{STAR-RIS}{STAR-RIS}{simultaneously transmitting and reflecting reconfigurable intelligent surface}
\newacronym{DoA}{DoA}{direction-of-arrival}
\newacronym{DD}{DD}{doubly-dispersive}
\newacronym{ODDM}{ODDM}{orthogonal delay-Doppler division multiplexing}
\newacronym{LoS}{LoS}{line-of-sight}
\newacronym{NLoS}{NLoS}{non-line-of-sight}
\newacronym{6G}{6G}{sixth generation}
\newacronym{MPDD}{MPDD}{metasurfaces-parameterized DD}
\newacronym{GaBP}{GaBP}{Gaussian Belief Propagation}
\newacronym{MSE}{MSE}{mean-squared-error}
\newacronym{sIC}{soft IC}{soft interference cancellation}
\newacronym{soft RG}{soft RG}{soft replica generation}
\newacronym{BG}{BG}{belief generation}
\newacronym{SGA}{SGA}{scalar Gaussian approximation}
\newacronym{CLT}{CLT}{central limit theorem}
\newacronym{PDF}{PDF}{probability density function}
\newacronym{QPSK}{QPSK}{quadrature phase-shift keying}
\newacronym{LMMSE}{LMMSE}{linear minimum mean square error}
\newacronym{SNR}{SNR}{signal-to-noise ratio}
\newacronym{QoS}{QoS}{quality of service}
\newacronym{CoV}{CoV}{calculus of variations}
\newacronym{CAPA}{CAPA}{continuous aperture array}
\newacronym{FCAPA}{FCAPA}{flexible continuous aperture array}
\newacronym{GL}{GL}{Gauss-Legendre}
\newacronym{DDC MIMO}{DDC MIMO}{DD continuous MIMO}
\newacronym{B5G}{B5G}{beyond fifth generation}
\newacronym{VR}{VR}{virtual reality}
\newacronym{XR}{XR}{extended reality}
\newacronym{ITN}{ITN}{intelligent traffic networks}
\newacronym{SAGIN}{SAGIN}{space-air-ground integrated network}
\newacronym{UAV}{UAV}{unmanned aerial vehicle}
\newacronym{MUSIC}{MUSIC}{Multiple Signal Classification}
\newacronym{ICC}{ICC}{integrated communication and computing}
\newacronym{SINR}{SINR}{signal-to-interference-plus-noise ratio}
\newacronym{WSR}{WSR}{weighted sum rate}
\newacronym{ARPU}{ARPU}{average rate per user}
\newacronym{BCD}{BCD}{block coordinate descent}
\newacronym{PDE}{PDE}{partial differential equation}
\newacronym{EL}{EL}{Euler-Lagrange}
\newacronym{TCA}{TCA}{tightly coupled array}
\newacronym{ELAA}{ELAA}{extremely large-aperture arrays}
\newacronym{LIS}{LIS}{large intelligent surface}
\newacronym{CSI}{CSI}{channel state information}
\newacronym{RIS}{RIS}{reconfigurable intelligent surface}

\begin{document}

\title{Flexible Continuous Aperture Arrays}

\author{Kuranage Roche Rayan Ranasinghe\textsuperscript{\orcidlink{0000-0002-6834-8877}},~\IEEEmembership{Graduate Student Member,~IEEE,}\,
Zhaolin Wang\textsuperscript{\orcidlink{0000-0003-4614-0175}},~\IEEEmembership{Member,~IEEE,} \\ 
Giuseppe Thadeu Freitas de Abreu\textsuperscript{\orcidlink{0000-0002-5018-8174}},~\IEEEmembership{Senior Member,~IEEE,}
and Emil Bj{\"o}rnson\textsuperscript{\orcidlink{0000-0002-5954-434X}},~\IEEEmembership{Fellow,~IEEE}
\thanks{K. R. R. Ranasinghe and G. T. F. de Abreu are with the School of Computer Science and Engineering, Constructor University (previously Jacobs University Bremen), Campus Ring 1, 28759 Bremen, Germany (emails: \{kranasinghe, gabreu\}@constructor.university).}
\thanks{Z. Wang is with the Department of Electrical and Electronic Engineering, The University of Hong Kong, Hong Kong (email: zhaolin.wang@hku.hk).}
\thanks{E. Bj{\"o}rnson is with the School of Electrical Engineering and Computer Science, KTH Royal Institute of Technology, Stockholm 16440, Sweden (email: emilbjo@kth.se).}
\vspace{-3ex}}



\maketitle

\begin{abstract}
A novel \ac{EM} structure termed \textit{\ac{FCAPA}} is proposed, which incorporates inherent surface flexibility into typical \ac{CAPA} systems, thereby enhancing the \ac{DoF} of \ac{MIMO} systems equipped with this technology.
By formulating and solving a downlink multi-user beamforming optimization problem to maximize the \ac{WSR} of the multiple users with \ac{FCAPA}, it is shown that the proposed structure outperforms typical \ac{CAPA} systems by a wide margin, with performance increasing with increasing morphability.

\end{abstract}

\begin{IEEEkeywords}
FCAPA, MIMO, CAPA, functional optimization, calculus of variations.
\end{IEEEkeywords}

\glsresetall

\vspace{-3ex}
\section{Introduction}

\IEEEPARstart{M}{ulti-antenna} technologies have been central to the evolution of wireless systems \cite{larsson2014massive, 10144733, dang2020should}. 
Array signal processing has enabled spatial multiplexing, interference suppression, and adaptive coverage shaping across various frequency bands, from sub-6 GHz to millimeter-wave. 
As wireless networks demand higher capacity and reliability, the industry has progressed from conventional sector antennas to massive \ac{MIMO} \cite{marzetta2010noncooperative}, \ac{ELAA} \cite{Bjornson2019, wang2024tutorial}, and holographic arrays with ultra-dense antenna deployment \cite{comsoc_news, deng2022reconfigurable, RanasingheTWC2025DDSIM}, enabling fine-grained manipulation of \ac{EM} wavefronts to improve communication performance. 
These trends point to a common direction: exploiting ever-larger antenna arrays with finer spatial control to unlock additional spatial \acp{DoF}, improve energy concentration, and tailor propagation to the environment.

This pursuit of fine-grained spatial control has advanced along two emerging, yet complementary, lines of research. 
The first focuses on continuous \ac{EM} control, formalized by \acp{CAPA} \cite{11095329}\footnotemark.
\Acp{CAPA} models the radiating aperture as a continuous current distribution rather than a finite set of discrete elements. 
In contrast to conventional discretized arrays, this approach treats beam synthesis as the shaping of a spatial \ac{EM} field over a surface, enabling precise pattern formation and providing deep insights into the impact of \ac{EM} properties---such as the near-field and polarization---on communications performance. 
The second line of research introduces mechanical and geometric control through flexible (or morphable) intelligent metasurfaces, formally known as \acp{FIM} \cite{AnTWC2025FIM, bai2022dynamically}.
By allowing a large programmable sheet to bend, stretch, or reconfigure its \ac{3D} shape while maintaining electronically tunable unit responses, these surfaces adapt their physical form to the radio environment or installation constraints. 
These two promising lines of development, one controlling the \ac{EM} field and the other, the physical geometry, suggest a natural unification.

\footnotetext{It is noteworthy that there also exists prior work on \acp{LIS} \cite{HuTSP2018,DardariJSAC2020} which consider densely populated antenna structures resembling a discretized \ac{CAPA}.
The fundamental difference between \acp{CAPA} and \acp{LIS} lies in their modeling approaches since \acp{LIS} uses discrete models to approximate \ac{EM} source currents, inevitably leading to discretization losses.}

\subsection{Prior Works}

\subsubsection{Studies on CAPAs}

Realizing \acp{CAPA} is a long-term and foundational goal in antenna design. 
In particular, Wheeler proposed the concept of the “current sheet" in the 1960s \cite{wheeler2003simple}, which is a theoretical plane that can support a continuous flow of \ac{EM} current, to understand the fundamental performance limits of phased arrays. 
This concept has served as a theoretical upper bound, inspiring researchers to develop practical antenna forms that approximate an optimal current sheet, such as \acp{TCA} \cite{munk2003finite, 8000624} and, more recently, holographic metasurfaces \cite{9886906, 9136592}. 
From a wireless communications perspective, the study of \acp{CAPA} is rooted in \ac{EM} information theory, which analyzes communication systems using fundamental \ac{EM} principles. 
Early studies focused on characterizing the ultimate performance bounds between continuous \ac{EM} volumes, such as the channel \acp{DoF} \cite{bucci1989degrees, miller2000communicating, dardari2020communicating, 9798854} and capacity \cite{jensen2008capacity, 8585146, wan2023mutual, 10012689}.
Recently, research efforts have shifted to developing novel signal processing techniques that address the unique challenges posed by the continuous, infinite-dimensional signal model of \acp{CAPA}. 
For instance, a wavenumber-domain discretization method was proposed to optimize the continuous signals in \cite{9906802, zhang2023pattern, 10612761}. 
This approach effectively transforms the functional optimization problem into a conventional discrete one, but at the cost of inevitable discretization loss and high computational complexity. 
As a remedy, the authors of \cite{WangTWC2025} proposed to solve the functional optimization problem directly by applying the \ac{CoV}, achieving improved performance with significantly reduced complexity. 
The \ac{CoV}-based method has since been successfully extended to solving the continuous signal optimization problem in various other cases \cite{wang2025beamformingdesigncontinuousaperture, OuyangTWC2025, ranasinghe2025doubly,ranasinghe2025lowcompCAPArec}.

\subsubsection{Studies on FIMs}

\Acp{FIM} have attracted significant attention, extending conventional reflective/transmissive metasurface technologies by adding shape and morphology control on top of per-element \ac{EM} tuning. 
Driven by advances in micro/nano-fabrication and flexible metamaterials, various \ac{FIM} implementations have been developed using flexible substrates that exhibit excellent \ac{EM} and mechanical properties \cite{kamali2016decoupling, qian2020deep, zhang2021photonic}. 
In particular, the authors of \cite{bai2022dynamically} and \cite{ni2022soft} proposed new \ac{FIM} types with programmable morphing among various surface shapes, enabling a dynamic response to the radio environment. 
Building on these advancements, the authors of \cite{AnTWC2025FIM} proposed exploiting \acp{FIM} to improve wireless communication performance, developing a joint optimization of transmit beamforming and \ac{FIM} shaping. 
More specifically, they investigated the capacity limits of \ac{FIM}-enabled \ac{MIMO} systems over frequency-flat fading channels by jointly optimizing the \ac{3D} surface geometries of both the transmit and receive \acp{FIM}, together with the transmit covariance matrix.
Simulation results revealed that \acp{FIM} can achieve up to a twofold increase in \ac{MIMO} capacity compared to traditional rigid array counterparts in certain configurations.
Additionally, \cite{AnTWC2025} explored \ac{FIM}-assisted multi-user downlink communications, where the goal was to minimize the transmit power at the \ac{BS} through a joint optimization of the transmit beamforming vectors and the \ac{FIM} surface morphology, subject to user \ac{QoS} constraints and the physical morphing limits of the \ac{FIM}.
The numerical analysis showed that such flexibility can yield a transmit power reduction of nearly $3$ dB relative to conventional rigid \ac{2D} arrays while maintaining equivalent throughput.
More recently, \cite{TVT_2025_Teng_Flexible} analyzed the influence of \ac{FIM} on wireless sensing under per-antenna power constraints, demonstrating a 3 dB enhancement in the total probing power at target locations by optimizing the transmitter-side \ac{FIM} surface shape.
Building on these advances, \cite{ranasinghe2025flexible} further extended prior studies by developing a new channel model that incorporates \acp{FIM} within doubly-dispersive propagation environments, enabling compatibility with advanced modulation formats such as \ac{OFDM}, \ac{OTFS}, and \ac{AFDM}.

\subsection{Motivations and Contributions}

Motivated by the complementary advantages of \acp{CAPA} and \acp{FIM}, this paper proposes the concept of the \ac{FCAPA} to realize their unification. 
\Acp{FCAPA} are defined as continuous radiating surfaces that provide fine-grained \ac{EM} controllability and whose aperture geometry can be dynamically reconfigured. 
Compared to conventional rigid \acp{CAPA}, bringing geometry into the aperture design enables the surface to not only conform to deployment constraints but also to exploit favorable structural forms. 
This geometric flexibility provides an additional dimension of control, enabling \ac{EM} field transmissions and shaping capabilities that are challenging or inefficient to achieve with electronic control alone. 
The result is an \ac{EM} surface capable of tailoring its radiation characteristics to the environment by controlling both “what the surface radiates” and “how the surface is shaped.” 
Building on these benefits, we formalize an \ac{FCAPA} signal and geometry model, develop a co-design framework that jointly optimizes aperture shape and continuous current distribution, and demonstrate the communication gains achievable when electronic and geometric controls are coordinated. 
The key contributions of this paper are summarized as follows:
\begin{itemize}
    \item We derive a mathematically rigorous model for a novel \ac{EM} structure termed \ac{FCAPA}, which unifies the continuous nature of \acp{CAPA} with the flexibility of \acp{FIM}.
    \item We formulate and solve a \ac{WSR} maximization problem with closed-form gradient expressions for the surface shape updates derived via the \ac{CoV}, envelope theorem and \ac{EL} conditions.
    \item The presented numerical analysis demonstrates major performance improvements with respect to both conventional \ac{CAPA} systems and typical \ac{FIM}-based structures.
\end{itemize}

\subsection{Organization and Notation}

\textit{Organization:} The rest of this paper is organized as follows. 
Section \ref{sec:Sys_Model} formalizes the mathematical model for an \ac{FCAPA} and formulates a downlink multi-user optimization problem.
Section \ref{sec:cov_el_sol} proposes a three stage solution to the non-convex optimization problem to consecutively optimize a set of auxiliary variables, the source currents and the surface shape of the \ac{FCAPA}.
Finally, Section \ref{sec:perf_analy} provides a detailed performance analysis for the proposed structure and Section \ref{sec:conclusion} concludes the manuscript.

\textit{Notation:} All scalars are represented by upper or lowercase letters, while column vectors and matrices are denoted by bold lowercase and uppercase letters, respectively.
The diagonal matrix constructed from vector $\mathbf{a}$ is denoted by diag($\mathbf{a}$), while $\mathbf{A}\trans$, $\mathbf{A}\herm$, $\mathbf{A}^{1/2}$, and $\mathbf{A}(i,j)$ denote the transpose, Hermitian, square root and the $(i,j)$-th element of a matrix $\mathbf{A}$, respectively.
The convolution and Kronecker product are respectively denoted by $*$ and $\otimes$, while $\mathbf{I}_N$ and $\mathbf{F}_N$ represent the $N\times N$ identity and the normalized $N$-point \ac{DFT} matrices, respectively.
The sinc function is expressed as $\text{sinc}(a) \triangleq \frac{\sin(\pi a)}{\pi a}$, and $j\triangleq\sqrt{-1}$ denotes the elementary complex number.
The Dirac delta function is denoted by $\delta(.)$.
The Lebesgue measure of a Euclidean subspace $\mathcal{S}$ is denoted by $|\mathcal{S}|$.
The absolute value and Euclidean norm are denoted by $|\cdot|$ and $||\cdot||$, respectively.
%


\section{System Model and Problem Formulation}
\label{sec:Sys_Model}

\begin{figure}[ht!]
    \centering
    \includegraphics[width=\columnwidth]{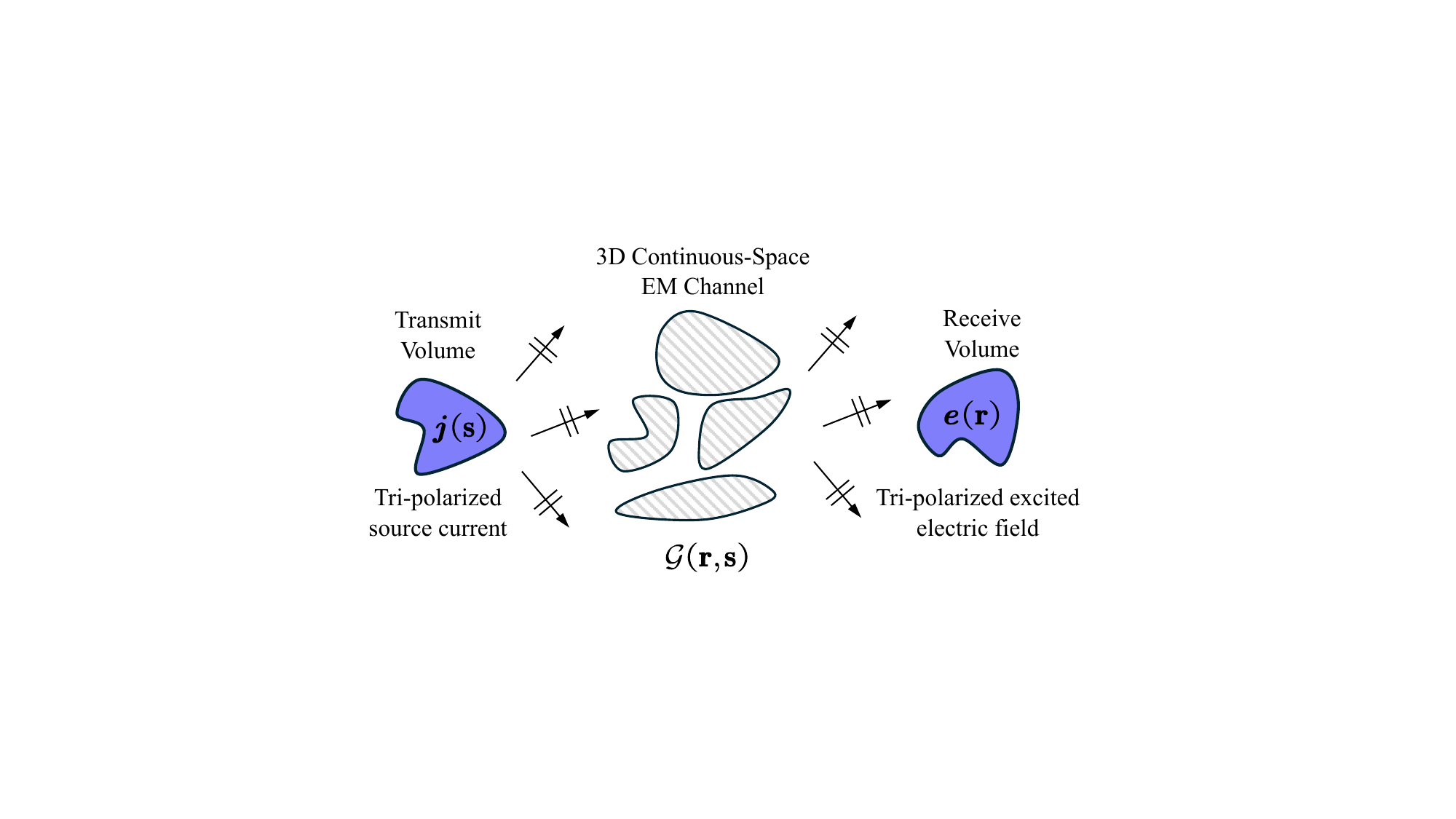}
    \caption{Illustration of a \ac{FCAPA}-based multi-user communications system with $K$ users, with an optional parametrization limit for the $y$-deformation defined by $[y_\text{min}, y_\text{max}]$.}
    \label{fig:FCAPA}
\end{figure}

We study an \ac{FCAPA}-based multi-user downlink communication system as illustrated in Fig. \ref{fig:FCAPA}.
The \ac{FCAPA} is mounted at a \ac{BS} to serve $K$ single-antenna users as shown.
Let us now leverage \ac{EM} principles to derive a complete signal model as follows.

\subsection{Transmit Signal Model}

Consider the \ac{3D} Euclidean space $\mathbb{R}^3$ with Cartesian coordinates $(x,y,z)$. 
We first define a standard \ac{CAPA} as a \ac{2D} region $\mathcal{D}$ embedded in the $x$–$z$ plane. 
This is represented by the parametrization
\begin{align}
\mathbf{d} : \mathcal{U} \subset \mathbb{R}^2 &\to \mathbb{R}^3, \\
(u,v) &\mapsto [\,u,\,0,\,v\,]\trans,
\end{align}
whose image is given by
\begin{align}
\mathcal{D} \;=\; \mathbf{d}(\mathcal{U})
= \Big\{ (x,y,z)^\top \in \mathbb{R}^3 : y=0,\; (x,z) \in \mathcal{U} \Big\}.
\end{align}

Next, we define a surface
\begin{align}
\mathbf{s} : \mathcal{U} &\to \mathbb{R}^3, \\
(u,v) &\mapsto [\,u,\,g(u,v),\,v\,]\trans,
\end{align}
whose image
\begin{align}
\mathcal{S} \!=\! \mathbf{s}(\mathcal{U})
\!=\! \Big\{ (x,y,z)^\top \in \mathbb{R}^3 : y = g(x,z), (x,z) \in \mathcal{U} \Big\}
\end{align}
represents a deformation of $\mathcal{D}$ in the $y$-direction that produces the physical model for the \ac{FCAPA}. 

It is noteworthy that $\mathcal{S} = \mathcal{D}$ when $g(u,v) = 0$; $i.e.,$ the model for \ac{FCAPA} becomes that of \ac{CAPA}.

Let $\mathbf{J}\big(\mathbf{s}(u,v),\omega\big) \in \mathbb{C}^{3\times 1}$ denote the Fourier transform of 
the source current density at the point
\begin{equation}
\label{eq:surface_shape_dep_s}
\mathbf{s}(u,v) = [\,u,\, g(u,v),\, v\,]\trans \in \mathcal{S},
\end{equation}
with $(u,v) \in \mathcal{U}$ explicitly parametrizing the surface $\mathcal{S}$, where $\omega = 2\pi f / c = 2\pi / \lambda$ denotes the angular frequency, $f$ is the signal frequency, and $\lambda$ is the signal wavelength.

In order to facilitate an initial setup with the \ac{FCAPA} model illustrated in Fig. \ref{fig:FCAPA}, we consider a narrowband\footnote{The extension to wideband multicarrier systems can be done by treating each subcarrier frequency separately.} single-carrier communication system, where the explicit dependence of the source current on $\omega$ can be omitted, such that the source current density can be expressed as $\mathbf{J}\big(\mathbf{s}(u,v)\big)$ and all quantities are normalized with respect to the bandwidth.
The integral of the source current density $\mathbf{J}\big(\mathbf{s}(u,v)\big)$ over the surface $\mathcal{S}$ can then be expressed in terms of the parameters $(u,v)$, as shall be described in the sequel.

To that end, first consider the partial derivatives of $\mathbf{s}(u,v)$ with respect to $u$ and $v$, which can be expressed as
\begin{subequations}
    \begin{equation}
        \partial_u \mathbf{s} \triangleq \frac{\partial \mathbf{s}(u,v)}{\partial u} = [1, \partial_u g, 0]\trans \in \mathbb{R}^{3 \times 1},
    \end{equation}
    \begin{equation}
        \partial_v \mathbf{s} \triangleq \frac{\partial \mathbf{s}(u,v)}{\partial v} = [0, \partial_v g, 1]\trans \in \mathbb{R}^{3 \times 1},
    \end{equation}
\end{subequations}
where we introduce the shorthand notation $\partial_u g \triangleq \frac{\partial g(u,v)}{\partial u}$ and $\partial_v g \triangleq \frac{\partial g(u,v)}{\partial v}$, for brevity.

Next, the magnitude of the normal vector, $i.e.,$ the area element on the surface $d\mathbf{s}$, can be computed as
\begin{equation}
    d \mathbf{s} \!=\! \| \partial_u \mathbf{s} \times \partial_v \mathbf{s} \| \, du \, dv = \underbrace{ \sqrt{1 + ( \partial_u g )^2 + ( \partial_v g )^2} }_{\triangleq \zeta(u,v)} \, du \, dv,
\end{equation}
where we define $\zeta(u,v) \triangleq \sqrt{1 + ( \partial_u g )^2 + ( \partial_v g )^2}$ for ease of notation.
Then, the integral $\mathbf{J}\big(\mathbf{s}(u,v)\big)$ over the surface $\mathcal{S}$ can be expressed as
\begin{equation}
\int_{\mathcal{S}} \!\!\mathbf{J}\big(\mathbf{s}(u,v)\big) \, d\mathbf{s} \!=\! \int_{\mathcal{U}} \!\!\mathbf{J}\big(\mathbf{s}(u,v)\big) \zeta(u,v) \, du \, dv.
\label{eq:param_map}
\end{equation}

Following \cite{WangTWC2025}, we consider the case of a vertically polarized transmitter, where the only excited component of the source current is in the $z$-direction.
Therefore, the source current can be expressed as
\begin{equation}
    \mathbf{J}\big(\mathbf{s}(u,v)\big) = J\big(\mathbf{s}(u,v)\big) \hat{\mathbf{u}}_z,
\end{equation}
where $\hat{\mathbf{u}}_z \in \mathbb{R}^{3 \times 1}$ is the unit vector along the $z$-axis.

To transmit $K$ information symbols to $K$ users, the scalar source current $J\big(\mathbf{s}(u,v)\big)$ can be cast as a linear superposition of $K$ information-bearing source currents, given by
\begin{equation}
    J\big(\mathbf{s}(u,v)\big) = \sum_{k=1}^K J_k\big(\mathbf{s}(u,v)\big) x_k,
\end{equation}
where $J_k\big(\mathbf{s}(u,v)\big) \in \mathbb{C}$ and $x_k \in \mathbb{C}$ represent the source current pattern and the communication symbol\footnote{In practice, one would first send a sequence of symbols using a pulse-shaping filter, and match that filter at the receiver before taking samples.
In addition, the Nyquist criterion has to be satisfied to avoid inter-symbol interference, which limits the number of symbols that can be transmitted per second.
However, since there are standard procedures to address these challenges, we adopt the discrete notation for brevity as commonly done in the \ac{SotA} \cite{WangTWC2025}.} for the $k$-th user, respectively. 

In addition, the communication symbols are assumed to be independent and have unit power, satisfying $\mathbb{E}[\mathbf{x}\mathbf{x}\herm] = \mathbf{I}_K$, where $\mathbf{x} \triangleq [x_1, \ldots, x_K]\trans \in \mathbb{C}^{K \times 1}$.

\subsection{EM Channel and Receive Signal Models}

Let $\mathbf{r}_k \in \mathbb{R}^{3\times 1}$ denote the position of the $k$-th user.
According to Maxwell’s equations and the relationship developed in \eqref{eq:param_map}, the electric field at $\mathbf{r}_k$ produced by the source current $\mathbf{J}\big(\mathbf{s}(u,v)\big)$ in a homogeneous medium is given by \cite{PoonTIT2005}
\begin{align}
\label{eq:ek_integral}
\mathbf{E}_k &= \!\!\int_{\mathcal{S}} \mathbf{G}\big(\mathbf{r}_k, \mathbf{s}(u,v)\big) \mathbf{J}\big(\mathbf{s}(u,v)\big) \, d\mathbf{s} \in \mathbb{C}^{3\times 1}  \\
&= \int_{\mathcal{U}} \!\!\mathbf{G}\big(\mathbf{r}_k, \mathbf{s}(u,v)\big) \mathbf{J}\big(\mathbf{s}(u,v)\big) \zeta(u,v) \, du \, dv. 
\nonumber
\end{align}
where in \ac{LoS} scenarios, $\mathbf{G}\big(\mathbf{r}_k, \mathbf{s}(u,v)\big) \in \mathbb{C}^{3\times 3}$ represents the Green's function. 

While $\mathbf{G}\big(\mathbf{r}_k, \mathbf{s}(u,v)\big)$ can be modeled as a stochastic process in various scattering environments, we focus on the \ac{LoS} channel in this work. 
It is important to note, however, that the proposed framework is not restricted to any particular channel model and maintains full generality.

In regions where the \ac{EM} field has settled into normal radiation, the Green’s function can be expressed as\footnote{The higher-order terms associated with reactive near-field effects are omitted since their impact on system performance is negligible.}
\begin{equation}
\mathbf{G}(\mathbf{r},\mathbf{s}) 
= -\frac{j\eta e^{-j \frac{2\pi}{\lambda} \|\mathbf{r}-\mathbf{s}\|}}{2 \lambda \|\mathbf{r}-\mathbf{s}\|} 
\Bigg( \mathbf{I}_3 - \frac{(\mathbf{r}-\mathbf{s})(\mathbf{r}-\mathbf{s})\trans}{\|\mathbf{r}-\mathbf{s}\|^2} \Bigg), \label{eq:green_function}
\end{equation}
where $\eta$ is the intrinsic impedance, and the inherent dependencies on $k$, $u$, and $v$ are omitted in $\mathbf{r}$ and $\mathbf{s}$, respectively.

Capturing the full \ac{3D} electric field $\mathbf{E}_k$ requires an ideal tri-polarized receiver at each $k$-th user, which is challenging in practice due to hardware and circuit limitations.
Consequently, we consider a more practical uni-polarized antenna for each $k$-th user with polarization direction $\hat{\mathbf{u}}_k \in \mathbb{R}^{3\times 1}$ satisfying $\|\hat{\mathbf{u}}_k\|=1$. 
In this setup, each $k$-th user only measures the component of $\mathbf{E}_k$ along $\hat{\mathbf{u}}_k$, resulting in a received field expressed by
\begin{align}
\label{eq:ek_noisy}
E_k &= \hat{\mathbf{u}}_k\trans \mathbf{E}_k + n_k  \\
&= \!\!\int_{\mathcal{U}} \!\!\hat{\mathbf{u}}_k\trans \mathbf{G}\big(\mathbf{r}_k, \mathbf{s}(u,v)\big) \mathbf{J}\big(\mathbf{s}(u,v)\big) \zeta(u,v) \, du \, dv + n_k, \nonumber
\end{align}
where $n_k \in \mathbb{C}$ denotes \ac{EM} noise factor, modeled as \ac{AWGN} with zero mean and variance $\sigma_k^2$, $i.e.,$ $n_k \sim \mathcal{CN}(0, \sigma_k^2)$ \cite{ZhangJSAC2023} which is the typical case when a passband filter is used in practice.

Separating the desired signal at a given $k$-th user from its interference yields the received field per $k$-th user as
\begin{align}
\label{eq:ek_total}
E_k &= \underbrace{\int_{\mathcal{U}} H_k\big(\mathbf{s}(u,v)\big) J_k\big(\mathbf{s}(u,v)\big) x_k \zeta(u,v) \, du \, dv }_{\text{desired signal}} \\
&\hspace{3ex}+ \underbrace{ \sum_{i\neq k}^{K} \int_{\mathcal{U}} H_i\big(\mathbf{s}(u,v)\big) J_i\big(\mathbf{s}(u,v)\big) x_i \zeta(u,v) \, du \, dv }_{\text{inter-user interference}} + n_k, \nonumber
\end{align}
where $H_k\big(\mathbf{s}(u,v)\big)$ represents the continuous \ac{EM} channel for the $k$-th user defined as
\begin{equation}
H_k\big(\mathbf{s}(u,v)\big) \triangleq \hat{\mathbf{u}}_k\trans \mathbf{G}\big(\mathbf{r}_k, \mathbf{s}(u,v)\big) \hat{\mathbf{u}}_z. 
\label{eq:continuous_channel}
\end{equation}

\subsection{Achievable Communication Rate}

To evaluate the achievable rate, the \ac{SINR} must be determined. 
Let $\varepsilon_k$ denote the absorption efficiency at the $k$-th user's receiving antenna. 
Then, leveraging \eqref{eq:ek_total}, the expected power received by the $k$-th user can be expressed as
\begin{equation}
\label{eq:received_power}
P_k = \mathbb{E} \Big[ \frac{\varepsilon_k}{2\eta} |E_k|^2 \Big] = \frac{\varepsilon_k}{2\eta}\left( \sum_{i=1}^{K} \left| a_i^{(k)} \right|^2 + \sigma_k^2 \right),
\end{equation}
where 
\begin{equation}
    a_i^{(k)} \!\triangleq \!\!\!\int_{\mathcal{U}} \!\!H_k\big(\mathbf{s}(u,v)\big)\,J_i\big(\mathbf{s}(u,v)\big)\,
\zeta(u,v)\; du\,dv,
\end{equation}
and we used the fact that $\mathbb{E}[\mathbf{x}\mathbf{x}\herm] = \mathbf{I}_K$.

Subsequently, the resulting \ac{SINR} for decoding the desired signal at a $k$-th user is given by
\begin{equation}
\gamma_k \;=\;
\frac{\big|a_k^{(k)}\big|^2}{\sum_{i=1,\,i\neq k}^{K} \big|a_i^{(k)}\big|^2 \;+\; \sigma_k^2 }.
\label{eq:SINR}
\end{equation}

Under the standard information-theoretic assumptions of perfect \ac{CSI}, Gaussian signaling, and interference treated as noise, the achievable rate for user 
$k$ can then be expressed as $\log_2(1 + \gamma_k)$.

\subsection{Problem Formulation}

For a set of $K$ users, the \ac{WSR} maximization problem can then be expressed as
\begin{align}
\underset{\big\{J_k\big(\mathbf{s}(u,v)\big)\big\}_{k=1}^K, \partial_u g, \partial_v g}{\text{maximize}} \quad & \sum_{k=1}^K \alpha_k \log_2(1 + \gamma_k), \label{eq:deformable_obj} \\
&\hspace{-23ex}\text{s.t.} \sum_{k=1}^K \int_{\mathcal{U}} \!\!\big|J_k\big(\mathbf{s}(u,v)\big)\big|^2\,
\zeta(u,v)\; du\,dv \le P_\mathrm{T}, \label{eq:deformable_cons}
\end{align}
where $\alpha_k$ is the weight specified for a given $k$-th user which can be determined according to the fairness and quality of service requirements, and $P_\mathrm{T}$ is the transmit power with units A$^2$, already incorporating the effects of the symbol rate. 

In addition, constraint \eqref{eq:deformable_cons} limits the transmit power of the \ac{FCAPA} transmitter.
Notice that the optimization problem in \eqref{eq:deformable_obj} is a non-convex functional programming problem where one has to jointly optimize a function and its derivatives simultaneously.
These types of problems can be solved via the \ac{CoV} technique \cite{Gelfand2000}.

\section{CoV-EL-based Solution}
\label{sec:cov_el_sol}

In this section, we derive a \ac{CoV}-based solution that leverages the envelope theorems and \ac{EL} conditions to directly optimize the source current patterns $\big\{J_k\big(\mathbf{s}(u,v)\big)\big\}_{k=1}^K$ and the surface shape of the \ac{FCAPA} via the derivatives $\partial_u g, \partial_v g$ in order to maximize the \ac{WSR}.

\subsection{Problem Reformulation}

To facilitate the optimization procedure, we first reformulate
problem \eqref{eq:deformable_obj} into an unconstrained optimization problem by
invoking the following lemmas.

\begin{lemma}[\textit{Equality Power Constraint}]
\label{lemma:equ_pow_const}
The optimal solution to problem \eqref{eq:deformable_obj} satisfies the
power constraint with equality, $i.e.,$
\begin{equation}
\sum_{k=1}^K \int_{\mathcal{U}} \!\!\big|J_k\big(\mathbf{s}(u,v)\big)\big|^2\,
\zeta(u,v)\; du\,dv = P_\mathrm{T}. \label{eq:lemma_eq_pow_constr}
\end{equation}

\begin{proof}
Let $\big\{\tilde{J}_k\big(\mathbf{s}(u,v)\big)\big\}_{k=1}^K$, $\partial_u \tilde{g}$, and $\partial_v \tilde{g}$ denote a set of feasible solutions to problem \eqref{eq:deformable_obj} that satisfies
\begin{equation}
\label{eq:proof_power_equi}
\tilde{P}_\mathrm{T} \!\triangleq\! \sum_{k=1}^K \int_{\mathcal{U}} \!\!\big|\tilde{J}_k\big(\mathbf{s}(u,v)\big)\big|^2\, \tilde{\zeta}(u,v)
\; du\,dv < P_\mathrm{T},
\end{equation}
where we intrinsically define $\tilde{\zeta}(u,v) \triangleq \sqrt{1 \!+\! (\partial_u \tilde{g})^2 \!+\! (\partial_v \tilde{g})^2}$.

Next, by defining a scaling factor $\rho_t \triangleq P_\mathrm{T}/\tilde{P}_\mathrm{T}$ and a scaled $k$-th solution $J_k\big(\mathbf{s}(u,v)\big) \triangleq \sqrt{\rho_t} \tilde{J}_k\big(\mathbf{s}(u,v)\big)$, it can easily be shown that the maximum objective in \eqref{eq:deformable_obj} achieved by the scaled $k$-th solution $J_k\big(\mathbf{s}(u,v)\big)$ must be higher than that achieved by the solution $\tilde{J}_k\big(\mathbf{s}(u,v)\big)$ since $\rho_t > 1$.

Additionally, it can also be shown that
\begin{align}
\label{eq:proof_power_equi_shown}
\sum_{k=1}^K &\int_{\mathcal{U}} \!\!\big|J_k\big(\mathbf{s}(u,v)\big)\big|^2\,
\zeta(u,v)\; du\,dv = \\
&\rho_t \sum_{k=1}^K \int_{\mathcal{U}} \!\!\big|\tilde{J}_k\big(\mathbf{s}(u,v)\big)\big|^2\, \tilde{\zeta}(u,v)\; du\,dv = \rho_t \tilde{P}_\mathrm{T} \!=\! P_\mathrm{T}.\nonumber
\end{align}

The results in \eqref{eq:proof_power_equi_shown} implies that for any feasible solution to \eqref{eq:deformable_obj}, there exists a solution that achieves a larger maximum objective with a corresponding power equality constraint. 
Note that we retain the same surface shape for the scaled solution, $i.e.,$ set $\partial_u g = \partial_u \tilde{g}$ and $\partial_v g = \partial_v \tilde{g}$, since the core argument---any feasible solution using less than full power $P_\mathrm{T}$ can be strictly improved by scaling only the current patterns $\{J_k\}$ while holding the shape (and its derivatives) fixed---is sufficient to establish that the global optimum must satisfy the power constraint with equality. 
Therefore, it is obvious that any deformation-based improvements are independently incremental to the power scaling gains.
The proof is therefore complete.
\end{proof}
\end{lemma}

\begin{lemma}[\textit{Unconstrained Equivalence Problem}]
\label{lemma:unconst_equi_prob}
Let $\big\{\bar{J}_k\big(\mathbf{s}(u,v)\big)\big\}_{k=1}^K$, $\partial_u \bar{g}$, and $\partial_v \bar{g}$ denote an optimal solution set to the functional maximization problem expressed as
\begin{equation}
\underset{\big\{J_k\big(\mathbf{s}(u,v)\big)\big\}_{k=1}^K, \partial_u g, \partial_v g}{\text{maximize}} \quad \sum_{k=1}^K \alpha_k \log_2(1 + \bar{\gamma}_k), 
\label{eq:deform_obj_lemma_unconstr}
\end{equation}
where
\begin{equation}
\bar{\gamma}_k \;=\;
\frac{\big|a_k^{(k)}\big|^2}{\sum_{i=1,\,i\neq k}^{K} \big|a_i^{(k)}\big|^2 \;+\; b_k }, 
\label{eq:deform_obj_lemma_unconstr_def_SINR}
\end{equation}
with $b_k$ given by
\begin{equation}
    b_k = \frac{\sigma_k^2}{P_\mathrm{T}} \sum_{i=1}^K \int_{\mathcal{U}} \!\!\big|J_i\big(\mathbf{s}(u,v)\big)\big|^2\,
\zeta(u,v)\; du\,dv.
\end{equation}

Then, an optimal solution to the original problem in \eqref{eq:deformable_obj} can be expressed as
\begin{equation}
J_k\big(\mathbf{s}(u,v)\big) = \sqrt{\frac{P_\mathrm{T}}{\bar{J}\big(\mathbf{s}(u,v)\big)}} \; \bar{J}_k\big(\mathbf{s}(u,v)\big), 
\label{eq:deformed_obj_uncon_sol}
\end{equation}
where
\begin{equation}
    \bar{J}\big(\mathbf{s}(u,v)\big) \!=\!\! \sum_{k=1}^K \int_{\mathcal{U}} \!\!\big|\bar{J}_k\big(\mathbf{s}(u,v)\big)\big|^2 \!\!\zeta(u,v)\; du\,dv.
\end{equation}

\begin{proof}  
The scaling in \eqref{eq:deformed_obj_uncon_sol} ensures that the equality power constraint in \eqref{eq:lemma_eq_pow_constr} is satisfied. 
Moreover, this transformation preserves the value of the objective function in the original problem \eqref{eq:deformable_obj}. 
Therefore, since $\bar{J}_k\big(\mathbf{s}(u,v)\big)$, $\partial_u \bar{g}$, and $\partial_v \bar{g}$ maximize the unconstrained problem in \eqref{eq:deform_obj_lemma_unconstr}, the corresponding $J_k\big(\mathbf{s}(u,v)\big)$, $\partial_u g$, and $\partial_v g$ in \eqref{eq:deformed_obj_uncon_sol} must also maximize the original constrained problem in \eqref{eq:deformable_obj}. 
\end{proof}
\end{lemma}

\begin{lemma}[\textit{Non-fractional Equivalence Problem}]
\label{lemma:non-frac_eqi_prob}
An equivalent formulation for the unconstrained problem in \eqref{eq:deform_obj_lemma_unconstr} is given by
\begin{align}
\underset{\big\{\mu_k, \lambda_k, J_k \big(\mathbf{s}(u,v)\big)\big\}_{k=1}^K, \partial_u g, \partial_v g}{\text{maximize}} \quad &\sum_{k=1}^K \alpha_k
\Bigg( 2\mu_k \Re \left\{\lambda_k^\ast a_k^{(k)} \right\} \nonumber \\
&- |\lambda_k|^2 \bigg(\sum_{i=1}^K \left|a_i^{(k)}\right|^2 + b_k \bigg) \Bigg), 
\label{eq:unonstrained_frac_equival}
\end{align}
where $\{\mu_k\}_{k=1}^K$ and $\{\lambda_k\}_{k=1}^K$ are auxiliary
variables.

\begin{proof}
This result follows directly from the quadratic transform
\cite[Theorem 2]{ShenTSP2018} and the Lagrangian dual transform
\cite[Theorem 3]{ShenTSP2018_Part2}. 
\end{proof}
    
\end{lemma}

Based on Lemmas \ref{lemma:equ_pow_const}, \ref{lemma:unconst_equi_prob}, and \ref{lemma:non-frac_eqi_prob}, the optimal solution of problem \eqref{eq:unonstrained_frac_equival} coincides with that of the original constrained problem in \eqref{eq:deformable_obj}.
In the next subsection, we propose a \ac{BCD}-\ac{CoV}-\ac{EL} algorithm to solve the problem stated in \eqref{eq:unonstrained_frac_equival}.

\subsection{BCD-CoV-EL Algorithm}

In problem \eqref{eq:unonstrained_frac_equival}, both the explicit constraints and the fractional structure in the objective function have been eliminated.
Consequently, the optimization variables are no longer coupled, which makes the problem more tractable.
This structure suggests the natural application of a \ac{BCD} approach, where each group of variables is optimized in turn, while the others are kept fixed.
For this purpose, we partition the variables into three distinct blocks: the auxiliary variables $\{\mu_k, \lambda_k\}_{k=1}^K$, the source currents $\big\{J_k\big(\mathbf{s}(u,v)\big)\big\}_{k=1}^K$, and the surface shape morphing parameters\footnote{There is also an intrinsic dependence on $g(u,v)$ (with the shorthand $g$ used hereafter) as well from equation \eqref{eq:surface_shape_dep_s}, but this will be discussed more explicitly in the morphing parameter optimization subproblem.} $\partial_u g$ and $\partial_v g$. 
The resulting subproblems for each block are discussed below.

\begin{figure*}
\setcounter{equation}{34}
\begin{align}
\label{eq:full_f}
f_k(J_k) &\triangleq 2 \Re \bigg\{A_k \int_{\mathcal{U}} H_k\big(\mathbf{s}(u,v)\big)\,J_k\big(\mathbf{s}(u,v)\big)\,
\zeta(u,v)\; du\,dv \bigg\} \\
&\hspace{3ex}- \sum_{i=1}^K \bigg( B_i \left| \int_{\mathcal{U}} H_k\big(\mathbf{s}(u,v)\big)\,J_i\big(\mathbf{s}(u,v)\big)
\zeta(u,v)\; du\,dv \right|^2 + C_i \int_{\mathcal{U}} \big|J_i\big(\mathbf{s}(u,v)\big)\big|^2
\zeta(u,v)\; du\,dv \bigg).
\nonumber
\end{align}

\setcounter{equation}{40}
\begin{align}
\label{eq:Phi_uv}
\Phi_k(\epsilon) \triangleq f_k(J_k + \epsilon U_k)
&=2\epsilon \Re\bigg\{
A_k \int_{\mathcal U} H_k^\ast\big(\mathbf{s}(u,v)\big)\,U_k^{\ast}(u,v) \, \zeta(u,v)\,du\,dv  \\
&- \sum_{i=1}^K \bigg( B_i
\int_{\mathcal U}\int_{\mathcal U} H_i\big(\mathbf{s}(u',v')\big)J_k\big(\mathbf{s}(u',v')\big) H_i^{\ast}\big(\mathbf{s}(u,v)\big) U_k^{\ast}(u,v)\,\zeta(u',v')\zeta(u,v)\,du'\,dv'\,du\,dv \nonumber \\
&+ C_i \int_{\mathcal U} J_k\big(\mathbf{s}(u,v)\big) U_k^{\ast}(u,v)\,\zeta(u,v)\,du\,dv \bigg)
\bigg\} \nonumber \\
&+ \epsilon^2 \sum_{i=1}^K\bigg(
B_i \Big|\int_{\mathcal U} H_i\big(\mathbf{s}(u,v)\big)U_k(u,v)\,\zeta(u,v)\,du\,dv\Big|^2
+ C_i \int_{\mathcal U} |U_k(u,v)|^2\,\zeta(u,v)\,du\,dv
\bigg) + D_k. \nonumber
\end{align}

\setcounter{equation}{31}
\hrulefill
\vspace{-3ex}
\end{figure*}

\subsubsection{Subproblem with respect to $\{\mu_k, \lambda_k\}_{k=1}^K$}

When $\big\{J_k\big(\mathbf{s}(u,v)\big)\big\}_{k=1}^K$, $\partial_u g$, $\partial_v g$ are fixed, problem \eqref{eq:unonstrained_frac_equival} reduces to a standard unconstrained optimization with respect to $\{\mu_k, \lambda_k\}_{k=1}^K$. 
Its optimal solution is given in \cite{ShenTSP2018,ShenTSP2018_Part2} as
\begin{align}
\mu_k &= \sqrt{1 + \bar{\gamma}_k}, \label{eq:muk_sol}\\
\lambda_k &= \frac{ \mu_k a_k^{(k)} }{ \sum_{i=1}^K \left|a_i^{(k)}\right|^2 + b_k }.
\label{eq:lambdak_sol}
\end{align}

\subsubsection{Subproblem with respect to $\big\{J_k\big(\mathbf{s}(u,v)\big)\big\}_{k=1}^K$}

Given a fixed set of $\{\mu_k, \lambda_k\}_{k=1}^K$ and $\partial_u g$, $\partial_v g$, the problem in \eqref{eq:unonstrained_frac_equival} can be expressed as
\begin{equation}
    \underset{\big\{J_k \big(\mathbf{s}(u,v)\big)\big\}_{k=1}^K}{\text{maximize}} \quad \sum_{k=1}^K f_k(J_k),
    \label{eq:cov_formulation}
\end{equation}
where the complete expanded form of $f_k(J_k)$ can be found in \eqref{eq:full_f} (top of next page), with the definitions $A_k \triangleq \alpha_k \mu_k \lambda_k^\ast$, $B_i \triangleq \alpha_i |\lambda_i|^2$ and $C_i \triangleq \frac{\alpha_i |\lambda_i|^2 \sigma_i^2}{P_\mathrm{T}}$.

It is evident that each of the three terms in $f_k(J_k)$ shown in \eqref{eq:full_f} is separable with respect to each set of functions $J_k$.
Hence, the solution can be obtained by independently determining the optimal set $J_k$ that maximizes the corresponding functional $f_k(J_k)$.
To tackle this type of functional optimization, the \ac{CoV} serves as a powerful and systematic approach and we follow the approach first described in \cite{WangTWC2025}.

Let us start with defining the fundamental lemma of \ac{CoV} under the space $\mathcal{U}$.

\begin{lemma}[\textit{Fundamental Lemma of \ac{CoV} in $\mathcal{U}$}]
\label{lemma:fund_lemma_cov}

Let $\mathcal{U}\subset\mathbb{R}^2$ be an open set with boundary $\partial\mathcal{U}$,
and let $\zeta(u,v)$ denote the surface Jacobian factor. 
For every smooth function $U(u,v)$ defined on $\mathcal{U}$, with the property
\setcounter{equation}{35}
\begin{equation}
U(u,v) = 0, \quad \forall (u,v) \in \partial \mathcal{U}, 
\end{equation}
if a continuous function $V(u,v)$ on $\mathcal{U}$ satisfies
\begin{equation}
\Re \left\{ \int_{\mathcal{U}} U^{\ast}(u,v)\,V(u,v)\,\zeta(u,v)\,du\,dv \right\} = 0,
\end{equation}
then it must follow that
\begin{equation}
V(u,v) = 0, \quad \forall (u,v) \in \mathcal{U}.
\end{equation}

\begin{proof}

This is a standard result from the \ac{CoV} literature and further details can be found in \cite{Gelfand2000}.
\end{proof}

\end{lemma}

Now, let us start by considering the functional $f_k(J_k)$ and its perturbed variation $J_k + \epsilon U_k$, where the variation $U_k(u,v)$ is an arbitrary smooth function satisfying
\begin{equation}
U_k(u,v)=0,\qquad \forall (u,v)\in\partial\mathcal{U}.
\end{equation}

We can now define the functional $\Phi_k(\epsilon)$ as
\begin{equation}
\Phi_k(\epsilon) \triangleq f_k(J_k + \epsilon U_k).
\end{equation}

Expanding $\Phi_k(\epsilon)$ using \eqref{eq:full_f} yields the form shown in \eqref{eq:Phi_uv} (top of the page), where $D_k$ is a constant independent of $\epsilon$.

Since $f_k(J_k)$ attains a local maximum at $J_k$, the functional $\Phi_k(\epsilon)$ has an extremum at $\epsilon=0$. Hence, we can write
\setcounter{equation}{41}
\begin{equation}
\label{eq:Phi_prime_zero}
\frac{d\Phi_k(\epsilon)}{d\epsilon}\Big|_{\epsilon=0}=0.
\end{equation}

Differentiating \eqref{eq:Phi_uv} with respect to $\epsilon$ and evaluating at $\epsilon=0$ yields
\begin{equation}
\Re\left\{ \int_{\mathcal U} U_k^{\ast}(u,v)\,V_k(u,v)\,\zeta(u,v)\,du\,dv \right\} = 0,
\label{eq:stationarity_uv}
\end{equation}
with
\begin{align}
\label{eq:Vk_uv}
V_k(u,v) &\triangleq A_k H_k^{\ast}\big(\mathbf{s}(u,v)\big) - \sum_{i=1}^K C_i J_k\big(\mathbf{s}(u,v)\big)  \\
&\hspace{-8ex}- \!\!\sum_{i=1}^K \!B_i H_i^{\ast}\!\big(\mathbf{s}(u,v)\big) \!\!\!\int_{\mathcal U} \!\!\!\!H_i\!\big(\mathbf{s}(u',v')\big)\! J_k\!\big(\mathbf{s}(u',v')\big)\zeta(u',v')du'dv'. \nonumber
\end{align}

The stationarity condition in \eqref{eq:stationarity_uv} holds for \emph{every} smooth variation $U_k$ vanishing on $\partial\mathcal U$. 
Leveraging Lemma \ref{lemma:fund_lemma_cov} with Jacobian $\zeta(u,v)$, it now follows that
\begin{equation}
\label{eq:Vk_zero_uv}
V_k(u,v)=0,\qquad \forall\,(u,v)\in\mathcal U.
\end{equation}
Substituting \eqref{eq:Vk_uv} into \eqref{eq:Vk_zero_uv} yields
\begin{equation}
\label{eq:Jk_final_uv}
J_k\big(\mathbf{s}(u,v)\big)
= \bar{A}_k H_k^{\ast}\big(\mathbf{s}(u,v)\big)
- \sum_{i=1}^K \bar{B}_i H_i^{\ast}\big(\mathbf{s}(u,v)\big) w_{k,i},
\end{equation}
where
\begin{equation}
    w_{k,i} \triangleq \int_{\mathcal U} H_i\big(\mathbf{s}(u',v')\big) J_k\big(\mathbf{s}(u',v')\big)\,\zeta(u',v')\,du'\,dv',
\end{equation}
with the normalized constants
\begin{equation}
\label{eq:bar_constants_uv}
\bar{A}_k \triangleq \frac{A_k}{\sum_{i=1}^K C_i},\qquad
\bar{B}_i \triangleq \frac{B_i}{\sum_{j=1}^K C_j}.
\end{equation}
Notice that \eqref{eq:Jk_final_uv} is a linear Fredholm integral equation of the second kind for $J_k\big(\mathbf{s}(u,v)\big)$.

It follows from \eqref{eq:Jk_final_uv} that once all coefficients $w_{k,i}$ for every $k$ and $i$ are determined, the function $J_k\big(\mathbf{s}(u,v)\big)$ can be readily computed, since $\bar{A}_k$, $\bar{B}_i$, and $H_k\big(\mathbf{s}(u,v)\big)$ are known for all $k,i$. 
Although $w_{k,i}$ depends on $J_k\big(\mathbf{s}(u,v)\big)$, it can be derived through the following steps.  

Multiplying both sides of \eqref{eq:Jk_final_uv} by $H_m\big(\mathbf{s}(u,v)\big)$, $\zeta(u,v)$ and integrating over $(u,v) \in \mathcal{U}$, we obtain
\begin{align}
&\int_{\mathcal{U}} H_m\big(\mathbf{s}(u,v)\big) J_k\big(\mathbf{s}(u,v)\big)\, {\zeta(u,v)} du \, dv \nonumber \\
&= 
\bar{A}_k \int_{\mathcal{U}} H_m\big(\mathbf{s}(u,v)\big) H_k^{\ast}\big(\mathbf{s}(u,v)\big)\, {\zeta(u,v)} du \, dv \nonumber \\
&- \sum_{i=1}^K w_{k,i} \bar{B}_i \int_{\mathcal{U}} H_m\big(\mathbf{s}(u,v)\big) H_i^{\ast}\big(\mathbf{s}(u,v)\big)\, {\zeta(u,v)} du \, dv.
\label{eq:fred_int_eq_sol}
\end{align}

Note that the left-hand side of \eqref{eq:fred_int_eq_sol} now corresponds to  $w_{k,m}$.
Then, by defining the channel correlation between users $i$ and $m$ as
\begin{equation}
q_{i,m} \triangleq \int_{\mathcal{U}} H_m\big(\mathbf{s}(u,v)\big) H_i^{\ast}\big(\mathbf{s}(u,v)\big)\, {\zeta(u,v)} du \, dv, \quad \forall i,m,
\label{eq:channel_corr}
\end{equation}
the expression in \eqref{eq:fred_int_eq_sol} can be rewritten as
\begin{equation}
w_{k,m} = \bar{A}_k q_{k,m} - \sum_{i=1}^K \bar{B}_i q_{i,m} w_{k,i}.
\label{eq:new_fred_sol}
\end{equation}

Next, let us define a matrix $\mathbf{W} \in \mathbb{C}^{K \times K}$ such that the entry in the $k$-th column and $i$-th row is given by $\mathbf{W}(k,i) = w_{k,i}$. 
Then, the set of equations in \eqref{eq:new_fred_sol} can be compactly represented in matrix form as
\begin{equation}
\label{eq:Fred_matrix_form}
\mathbf{W} = \mathbf{Q}\mathbf{A} - \mathbf{Q}\mathbf{B}\mathbf{W}
\implies
(\mathbf{I}_K + \mathbf{Q}\mathbf{B})\mathbf{W} = \mathbf{Q}\mathbf{A},
\end{equation}

where
\begin{align}
\mathbf{A} &= \operatorname{diag}\{\bar{A}_1, \cdots, \bar{A}_K\} \in \mathbb{C}^{K \times K}, \label{eq:A_def}\\
\mathbf{B} &= \operatorname{diag}\{\bar{B}_1, \cdots, \bar{B}_K\} \in \mathbb{R}_+^{K \times K}, \label{eq:B_def}\\
\mathbf{Q} &= [\mathbf{q}_1, \cdots, \mathbf{q}_K] \in \mathbb{S}_+^{K \times K}, \label{eq:Q_def}\\
\mathbf{q}_k &= [q_{k,1}, \cdots, q_{k,K}]\trans \in \mathbb{C}^{K \times 1}. \label{eq:q_def}
\end{align}

Since $\mathbf{Q}$ is positive semidefinite and $\mathbf{B}$ is a diagonal matrix with positive elements, the matrix $(\mathbf{I}_K + \mathbf{Q}\mathbf{B})$ is guaranteed to be invertible. 
Hence, the matrix $\mathbf{W}$ can be explicitly obtained as
\begin{equation}
\mathbf{W} = (\mathbf{I}_K + \mathbf{Q}\mathbf{B})^{-1} \mathbf{Q}\mathbf{A}.
\label{eq:final_w}
\end{equation}

Finally, by substituting the elements $w_{k,i}$ from $\mathbf{W}$ into \eqref{eq:Jk_final_uv}, the optimal functional $J_k\big(\mathbf{s}(u,v)\big)$ that maximizes the expression in \eqref{eq:cov_formulation} can be calculated.

\begin{figure*}
\setcounter{equation}{65}
\begin{equation}
\mathcal{F}_k(u,v,g,\partial_u g, \partial_v g) \triangleq 2 \Re \big\{A_k H_k(u,v) J_k^\star(\cdot;g)
\zeta(u,v) \big\}- \sum_{i=1}^K \big( B_i \big| {I_{k,i}[g]} \big|^2 + C_i \big|J_i^\star(\cdot;g)\big|^2
\zeta(u,v) \big).
\label{eq:local_integrand}
\end{equation}

\setcounter{equation}{75}
\begin{align}
\label{eq:euler_lagrange_final}
G(u,v) \triangleq \sum_{k=1}^{K} 
\Bigg[
&2\,\Re\!\bigg\{A_k \frac{H_k(u,v)}{\partial g} J_k^\star(\cdot;g) \zeta(u,v)\bigg\} - 2\! \sum_{i=1}^{K} \!B_i \Re\!\bigg\{\! {I_{k,i}^*[g] \int_{\mathcal U}\!\! \frac{H_k(u',v') }{\partial g} J_i^\star(\cdot;g);g\big)\,\zeta(u',v')\,du'\,dv'} \!\bigg\}  \\
&- \partial_u \Bigg( \frac{\partial_u g}{\zeta} \!\bigg( \underbrace{ 2\,\Re\!\{A_k H_k(u,v) J_k^\star(\cdot;g)\} - {2} \sum_{i=1}^{K} \Big( B_i\,{\Re\!\{ I_{k,i}^*[g] H_k(u,v) J_i^\star(\cdot;g)\}} + C_i\,|J_i^\star(\cdot;g)|^2 \Big)}_{\triangleq G_d(u,v)} \bigg) \Bigg) \nonumber \\
&- \partial_v \Bigg( \frac{\partial_v g}{\zeta} \!\bigg( \overbrace{ 2\,\Re\!\{A_k H_k(u,v) J_k^\star(\cdot;g)\} - {2} \sum_{i=1}^{K} \Big( B_i\,{\Re\!\{ I_{k,i}^*[g] H_k(u,v) J_i^\star(\cdot;g)\}} + C_i\,|J_i^\star(\cdot;g)|^2 \Big)} \bigg) \Bigg) 
\Bigg] = 0. \nonumber
\end{align}

\setcounter{equation}{56}
\hrulefill
\vspace{-3ex}
\end{figure*}

\subsubsection{Subproblem with respect to $\partial_u g$, $\partial_v g$}

Given a fixed set of auxiliary variables $\{\mu_k, \lambda_k\}_{k=1}^K$ and source currents $\big\{J_k\big(\mathbf{s}(u,v)\big)\big\}_{k=1}^K$, the problem in \eqref{eq:unonstrained_frac_equival} can be expressed as
\setcounter{equation}{57}
\begin{equation}
    \underset{g, \partial_u g, \partial_v g}{\text{maximize}} \quad \sum_{k=1}^K f_k(J_k^\star\big(\mathbf{s}(u,v);g\big), g, \partial_u g, \partial_v g),
    \label{eq:shape_formulation}
\end{equation}
where we explicitly indicate the dependence of $f_k$ on $g$, $\partial_u g$, and $\partial_v g$ through the surface parametrization $\mathbf{s}(u,v)$ in \eqref{eq:surface_shape_dep_s} and the Jacobian factor $\zeta(u,v)$ in \eqref{eq:full_f}.

Here, $f_k(J_k^\star\big(\mathbf{s}(u,v);g\big), g, \partial_u g, \partial_v g)$ is identical to $f_k(J_k)$ in \eqref{eq:full_f}, except that we now define the optimal current per the $k$-th user for a fixed $g$ as $J_k^\star\big(\mathbf{s}(u,v);g\big)$, which is the solution to the subproblem in \eqref{eq:cov_formulation} for given $g$. 

To solve problem \eqref{eq:shape_formulation}, we employ the \ac{CoV} method combined with the envelope theorem \cite{envelope_theorem}, which allows us to compute the variation of the objective functional without needing to evaluate the implicit dependence $\partial J_k^\star\big(\mathbf{s}(u,v);g\big)/\partial g$.

For conciseness, let us also define the shorthand notation $H_k(u,v) \equiv H_k\big(\mathbf{s}(u,v)\big)$ and $J_k^\star(\cdot;g) \equiv J_k^\star\big(\mathbf{s}(u,v);g\big)$, giving us the reduced objective functional
\begin{equation}
    \mathcal{J}[g] \triangleq \sum_{k=1}^{K} f_k\big(J_k^\star(\cdot;g), g\big),
    \label{eq:J_reduced_form}
\end{equation}
where the dependence of $f_k$ on $\partial_u g$ and $\partial_v g$ is implicit through $g$.

\begin{remark}
    We assume that for each fixed $g$, the functional $f_k\big(J, g\big)$ is twice continuously differentiable in $\big(J, g\big)$ and strictly concave in $J$.
    Then, the admissible $J_k$ forms a convex set and the maximizer $J_k^\star(\cdot;g)$ is unique and an interior point.
    Under these conditions, the envelope theorem applies and differentiation under the integral sign is permitted.
    The aforementioned assumptions hold for most analytical \ac{EM} optimization problems since $f_k\big(J, g\big)$ is usually quadratic or bilinear in $J$, and potential limitations only occur if $J_k$ is constrained in magnitude or phase (for $e.g.,$ if subjected to constant-modulus or quantized currents) or $f_k\big(J, g\big)$ has non-quadratic penalties (for $e.g.,$ from sparsity constraints or nonlinear mutual coupling).
    These cases will then have to be addressed on a case-by-case basis with solutions derived accordingly.
\end{remark}

Let $\eta(u,v)$ be an admissible perturbation of $g(u,v)$ such that
\begin{equation}
    g_\epsilon(u,v) = g(u,v) + \epsilon\,\eta(u,v), \qquad \epsilon \in \mathbb{R}.
\end{equation}
Then, the first variation of the functional $\mathcal{J}[g]$ can be obtained via its Gateaux derivative in the direction $\eta$ as
\begin{equation}
    \delta \mathcal{J}[g;\eta]
    \;\triangleq\;
    \left.
    \frac{d}{d\epsilon}
    \mathcal{J}[g+\epsilon\eta]
    \right|_{\epsilon=0},
    \label{eq:Gateaux_def_uv}
\end{equation}
where we drop the implicit dependence on $u,v$ for brevity. 

Next, applying the chain rule for functionals \cite[page 37, eq.~(6)]{greiner1996field} to \eqref{eq:J_reduced_form} yields
\begin{align}
    \delta \mathcal{J}[g;\eta]
    &=
    \sum_{k=1}^{K}
    \left.
    \frac{d}{d\epsilon}
    f_k\!\big(J_k^\star(\cdot;g_\epsilon),\, g_\epsilon\big)
    \right|_{\epsilon=0}
    \nonumber\\[3pt]
    &=
    \sum_{k=1}^{K}
    \Bigg(
        \underbrace{
        \left.\frac{\partial f_k}{\partial g}\right|_{J_k^\star,g}[\eta]
        }_{\text{explicit dependence w.r.t. $g$}}  \nonumber \\
        &\hspace{3ex}+
        \underbrace{
        \left\langle
        \left.\frac{\delta f_k}{\delta J_k}\right|_{J_k^\star,g},
        \left.
        \frac{d J_k^\star(\cdot;g_\epsilon)}{d\epsilon}
        \right|_{\epsilon=0}
        \right\rangle
        }_{\text{implicit dependence w.r.t. $g$ via } J_k^\star}
    \Bigg),
    \label{eq:Gateaux_expanded}
\end{align}
where $\langle \cdot , \cdot \rangle$ denotes the $L^2(\mathcal{U})$ inner product with respect to $du\,dv$.

Since $J_k^\star(\cdot;g_\epsilon)$ is the maximizer of $f_k\big(J_k^\star(\cdot;g), g\big)$ for each fixed $g$, its first-order optimality condition (stationarity) implies
\begin{equation}
    \left.
    \frac{\delta f_k}{\delta J_k}
    \right|_{J_k^\star,g}
    = 0,
    \qquad \forall k,
    \label{eq:inner_stationarity}
\end{equation}
and therefore the second (implicit) term in \eqref{eq:Gateaux_expanded} vanishes.  
This is the envelope theorem \cite{envelope_theorem}, yielding
\begin{equation}
    \delta\mathcal{J}[g;\eta]
    \;=\;
    \sum_{k=1}^{K}
    \left.
    \frac{\partial f_k}{\partial g}
    \right|_{J_k^\star,g}
    [\eta].
    \label{eq:envelope_applied_final}
\end{equation}
Introducing the \textit{local integrand} $\mathcal{F}_k(u,v,g,\partial_u g, \partial_v g)$ gives
\begin{equation}
    f_k\big(J_k^\star(\cdot;g), g\big) \triangleq \int_{\mathcal{U}} \mathcal{F}_k(u,v,g,\partial_u g, \partial_v g)\; du\,dv,
    \label{eq:fk_integrand_form}
\end{equation}
with $\mathcal{F}_k(u,v,g,\partial_u g, \partial_v g)$ defined in \eqref{eq:local_integrand} on the top of the page, {where we have implicitly defined the non-local term
\setcounter{equation}{66}
\begin{equation}
    I_{k,i}[g] \triangleq \int_{\mathcal U} H_k(u',v') J_i^\star(\cdot;g);g\big)\,\zeta(u',v')\,du'\,dv',
    \label{eq:Iik_def}
\end{equation}
which only depends on the inner variables $u'$ and $v'$.
}

Now, the first variation of $f_k\big(J_k^\star(\cdot;g), g\big)$ in the direction $\eta$ can be expressed as
\begin{equation}
\delta \mathcal{J}[g;\eta] \!=\! \sum_{k=1}^{K} \int_{\mathcal{U}} \bigg[ \frac{\partial \mathcal{F}_k}{\partial g} \, \eta + \frac{\partial \mathcal{F}_k}{\partial (\partial_u g)} \, \partial_u \eta + \frac{\partial \mathcal{F}_k}{\partial (\partial_v g)} \, \partial_v \eta
\bigg] du\,dv.
\end{equation}

Next, applying integration by parts (assuming $\eta=0$ on the boundary $\partial \mathcal{U}$) moves the derivatives off $\eta$, yielding
\begin{subequations}
\begin{equation}
\int_{\mathcal{U}} \frac{\partial \mathcal{F}_k}{\partial (\partial_u g)} \partial_u \eta \, du\,dv
= - \int_{\mathcal{U}} \partial_u \left( \frac{\partial \mathcal{F}_k}{\partial (\partial_u g)} \right) \eta \, du\,dv, 
\end{equation}
\begin{equation}
\int_{\mathcal{U}} \frac{\partial \mathcal{F}_k}{\partial (\partial_v g)} \partial_v \eta \, du\,dv
= - \int_{\mathcal{U}} \partial_v \left( \frac{\partial \mathcal{F}_k}{\partial (\partial_v g)} \right) \eta \, du\,dv.
\end{equation}
\end{subequations}

Since $\eta$ is arbitrary in the interior of $\mathcal{U}$, the Euler–Lagrange condition for the stationarity of $\mathcal{J}$ is
\begin{equation}
\sum_{k=1}^{K} 
\bigg[
\frac{\partial \mathcal{F}_k}{\partial g}
- \partial_u \left( \frac{\partial \mathcal{F}_k}{\partial (\partial_u g)} \right)
- \partial_v \left( \frac{\partial \mathcal{F}_k}{\partial (\partial_v g)} \right)
\bigg] \!=\! 0, \forall (u,v)\in\mathcal{U}.
\label{eq:euler_lagrange_general_form}
\end{equation}

Equation \eqref{eq:euler_lagrange_general_form} is now a nonlinear second order \ac{PDE} for the optimal surface shape $g(u,v)$\footnote{Strictly speaking, \eqref{eq:euler_lagrange_general_form} falls into a category of integro-\acp{PDE} due to the presence of  $I_{k,i}[g]$.
However, since this term can be precomputed in advance, we hereafter refer to  \eqref{eq:euler_lagrange_general_form} as a regular \ac{PDE}.}.
To obtain the explicit form of the \ac{EL} in \eqref{eq:euler_lagrange_general_form}, we need to compute the partial derivatives of $\mathcal{F}_k$ with respect to $g$, $\partial_u g$, and $\partial_v g$.

First, for the $\zeta$ dependence we have
\begin{equation}
\frac{\partial \zeta}{\partial (\partial_u g)} = \frac{\partial_u g}{\zeta}, \qquad
\frac{\partial \zeta}{\partial (\partial_v g)} = \frac{\partial_v g}{\zeta}, \qquad
\frac{\partial \zeta}{\partial g} = 0,
\end{equation}
so that for any term of the form $R(u,v)\, \zeta(u,v)$ in $\mathcal{F}_k$, we can obtain
\begin{subequations}
\begin{equation}
\frac{\partial}{\partial (\partial_u g)} [ R \zeta] = R \frac{\partial_u g}{\zeta},
\end{equation}
\begin{equation} 
\frac{\partial}{\partial (\partial_v g)} [ R \zeta] = R \frac{\partial_v g}{\zeta}, 
\end{equation}
\begin{equation}
\frac{\partial}{\partial g} [ R \zeta] = \frac{\partial R}{\partial g} \, \zeta.
\end{equation}
\end{subequations}

\begin{figure*}
\setcounter{equation}{77}

\begin{align}
    \label{eq:rho_simp}
    \rho(\mathbf{W}) &\triangleq \sum_{k=1}^K \int_{\mathcal{U}} \!\!\big|J_k\big(\mathbf{s}(u,v)\big)\big|^2\, \zeta(u,v)\; du\,dv = \sum_{k=1}^K \int_{\mathcal{U}} \bigg| \bar{A}_k H_k^{\ast}\big(\mathbf{s}(u,v)\big) - \sum_{i=1}^K \bar{B}_i H_i^{\ast}\big(\mathbf{s}(u,v)\big) w_{k,i} \bigg|^2\,\zeta(u,v)\; du\,dv  \\
    &\hspace{-6ex}= \sum_{k=1}^K \Bigg( \int_{\mathcal{U}} \bar{A}_k H_k^{\ast}\!\big(\mathbf{s}(u,v)\big)\, H_k\!\big(\mathbf{s}(u,v)\big)\, \bar{A}_k^{\ast} \zeta(u,v)\; du\,dv - \int_{\mathcal{U}} \bar{A}_k H_k^{\ast}\!\big(\mathbf{s}(u,v)\big)
    \sum_{i=1}^K w_{k,i}^{\ast} H_i\!\big(\mathbf{s}(u,v)\big)\, \bar{B}_i^{\ast} \zeta(u,v)\; du\,dv \nonumber\\
    &\hspace{-6ex}- \int_{\mathcal{U}} \sum_{i=1}^K \bar{B}_i H_i^{\ast}\!\big(\mathbf{s}(u,v)\big)\, w_{k,i}\,
    H_k\!\big(\mathbf{s}(u,v)\big)\, \bar{A}_k^{\ast} \zeta(u,v)\; du\,dv + \int_{\mathcal{U}} \sum_{i=1}^K \sum_{j=1}^K 
    \bar{B}_i H_i^{\ast}\!\big(\mathbf{s}(u,v)\big)\, w_{k,i}\, w_{k,j}^{\ast}\,
    H_j\!\big(\mathbf{s}(u,v)\big)\, \bar{B}_j^{\ast} \zeta(u,v)\; du\,dv \Bigg)   \nonumber \\
    &\hspace{-6ex}= \sum_{k=1}^K \!\Bigg(  \!|\bar{A}_k|^2 q_{k,k} \!-\! 
    \sum_{i=1}^K 2\Re\{ \bar{A}_k \bar{B}_i^{\ast} w_{k,i}^{\ast} q_{k,i}  \} \!+\! \sum_{i=1}^K \sum_{j=1}^K w_{k,i}\, w_{k,j}^{\ast}\, \bar{B}_i \bar{B}_j^{\ast} q_{i,j} \!\Bigg) = \text{tr}\big( \mathbf{A}\herm \mathbf{Q} \mathbf{A} - 2\Re\{ \mathbf{A} \mathbf{W}\herm \mathbf{B}\herm \mathbf{Q} \} + \mathbf{W}\herm \mathbf{B}\herm \mathbf{Q} \mathbf{B} \mathbf{W} \big).  \nonumber
\end{align}

\setcounter{equation}{72}
\hrulefill
\vspace{-3ex}
\end{figure*}

Let us once again leverage the envelope theorem \cite{envelope_theorem} to avoid computing the implicit dependence of $J_k^\star(\cdot;g)$ on $g$ when calculating the explicit partial derivatives of $\mathcal{F}_k$.

Applying the rules above to the surface gradient terms in \eqref{eq:local_integrand} yields
\begin{subequations}
\label{eq:F_partials}
\begin{align}
\frac{\partial \mathcal{F}_k}{\partial (\partial_u g)}
&= \frac{\partial_u g}{\zeta} \!\bigg( 2\,\Re\!\{A_k H_k(u,v) J_k^\star(\cdot;g)\} \\
&\hspace{-5ex}- {2} \sum_{i=1}^{K} \Big( B_i\,{\Re\!\{ I_{k,i}^*[g] H_k(u,v) J_i^\star(\cdot;g)\}} + C_i\,|J_i^\star(\cdot;g)|^2 \Big)\bigg), \nonumber
\end{align}
\begin{align}
\frac{\partial \mathcal{F}_k}{\partial (\partial_v g)}
&= \frac{\partial_v g}{\zeta} \!\bigg( 2\,\Re\!\{A_k H_k(u,v) J_k^\star(\cdot;g)\} \\
&\hspace{-5ex}- {2} \sum_{i=1}^{K} \Big( B_i\,{\Re\!\{ I_{k,i}^*[g] H_k(u,v) J_i^\star(\cdot;g)\}} + C_i\,|J_i^\star(\cdot;g)|^2 \Big)\bigg), \nonumber
\end{align}
\end{subequations}

Finally, the explicit partial derivative with respect to $g$ can be computed as
\begin{align}
    \label{eq:F_g_partial}
\frac{\partial \mathcal{F}_k}{\partial g}
&=  2\,\Re\!\bigg\{A_k \frac{H_k(u,v)}{\partial g} J_k^\star(\cdot;g) \zeta(u,v)\bigg\} \\
&\hspace{-5ex}- 2\! \sum_{i=1}^{K} \!B_i \Re\!\bigg\{\! {I_{k,i}^*[g] \int_{\mathcal U}\!\! \frac{H_k(u',v') }{\partial g} J_i^\star(\cdot;g);g\big)\,\zeta(u',v')\,du'\,dv'} \!\bigg\}. \nonumber
\end{align}

If, in addition, the local integrand $\mathcal{F}_k$
depends on $g$ only through its derivatives $\partial_u g$ and $\partial_v g$, and $H_k(u,v)$ has no explicit dependence on $g$, then
\begin{equation}
\frac{\partial \mathcal{F}_k}{\partial g} = 0.
\label{eq:F_g_zero}
\end{equation}

One could also reasonably approximate $\frac{\partial \mathcal{F}_k}{\partial g} \approx 0$ in far-field conditions where the deformations of $g(u,v)$ are small compared to the user distances $R_y^\text{min}$ and $R_y^\text{max}$. 

We can now substitute \eqref{eq:F_partials} and \eqref{eq:F_g_partial} into the \ac{EL} equation in \eqref{eq:euler_lagrange_general_form} to obtain the explicit \ac{PDE} for the optimal surface shape $g(u,v)$, given in \eqref{eq:euler_lagrange_final} on top of the previous page.
Notice that we also explicitly define the entire equation \eqref{eq:euler_lagrange_final} into a gradient field $G(u,v)$, which is the \ac{EL} residual.

Then, an ascent scheme to update an arbitrary surface shape $g(u,v)$ at each $i$-th iteration can be expressed as
\setcounter{equation}{76}
\begin{subequations}
\label{eq:Ful_gradeint_update}
\begin{align}
g^{(i+1)}(u,v) &\leftarrow \mathrm{proj}_{[y_\text{min},y_\text{max}]} \bigg( g^{(i)}(u,v) \\
&\hspace{-8ex}+ \nu_i \, G \Big(g^{(i)}(u,v), \zeta^{(i)}(u,v), \partial_u^2 g^{(i)}(u,v), \partial_v^2 g^{(i)}(u,v)\Big) \bigg), \nonumber
\label{eq:g_update}
\end{align}
\begin{equation}
    \partial_u g^{(i+1)}(u,v) \leftarrow \frac{\partial}{\partial u} \Big( g^{(i+1)}(u,v) \Big),
\end{equation}
\begin{equation}
    \partial_v g^{(i+1)}(u,v) \leftarrow \frac{\partial}{\partial v} \Big( g^{(i+1)}(u,v) \Big),
\end{equation}
\begin{equation}
    \zeta^{(i+1)}(u,v) \leftarrow \sqrt{1 \!+\! \big(\partial_u g^{(i+1)}(u,v)\big)^2 \!+\! \big(\partial_v g^{(i+1)}(u,v)\big)^2},
\end{equation}
\begin{equation}
    \partial_u^2 g^{(i+1)}(u,v) \leftarrow \frac{\partial}{\partial u} \Big( \partial_u g^{(i+1)}(u,v) \Big),
\end{equation}
\begin{equation}
    \partial_v^2 g^{(i+1)}(u,v) \leftarrow \frac{\partial}{\partial v} \Big( \partial_v g^{(i+1)}(u,v) \Big),
\end{equation}
with step-size $\nu_i > 0$ chosen by an Armijo line-search, and $\mathrm{proj}_{[y_\text{min},y_\text{max}]} (\cdot)$ denotes an intrinsic optional projection within a fixed boundary defined by $y_\text{min}$ and $y_\text{max}$ as also visualized in Fig. \ref{fig:FCAPA}.
\end{subequations}

While the above gradient ascent procedure is sub-optimal due to convergence to local maxima, since $J_k^\star(\cdot;g)$ is optimal for a fixed $g$ according to \cite{WangTWC2025}, it remains sufficient to show incremental gains due to the flexibility component introduced. 

Let us realize that while the auxiliary variables $\mu_k$ and $\lambda_k$ in \eqref{eq:muk_sol} and \eqref{eq:lambdak_sol} are given in a closed-form, they still involve the computations of integrals.
In order to resolve this matter and propose a low-complexity equivalent, notice that all the integrals in \eqref{eq:muk_sol} and \eqref{eq:lambdak_sol} are functions of $w_{k,m}$, except for the total power integral $\sum_{k=1}^K \int_{\mathcal{U}} \!\!\big|J_k\big(\mathbf{s}(u,v)\big)\big|^2\,\zeta(u,v)\; du\,dv$.

To resolve this, leveraging equation \eqref{eq:Jk_final_uv} yields \eqref{eq:rho_simp} on the top of the page, which only involves matrix operations.
Correspondingly, \eqref{eq:muk_sol} and \eqref{eq:lambdak_sol} can be expressed as
\vspace{-1ex}
\setcounter{equation}{78}
\begin{align}
\vspace{-1ex}
\mu_k(\mathbf{W}) &= \sqrt{1 + \frac{w_{k,k}}{ \sum_{i=1,\,i\neq k}^{K} \big|w_{i,k}\big|^2 \;+\; \frac{\sigma_k^2}{P_\mathrm{T}} \rho(\mathbf{W}) } }, \label{eq:muk_sol_W}\\
\lambda_k(\mathbf{W}) &= \frac{ \mu_k(\mathbf{W}) w_{k,k} }{ \sum_{i=1}^K \left|w_{i,k}\right|^2 + \frac{\sigma_k^2}{P_\mathrm{T}} \rho(\mathbf{W}) }.
\label{eq:lambdak_sol_W}
\end{align} 

Next, in order to update the values in $\mathbf{W}$ at each iteration, we have to compute the channel correlation $\mathbf{Q}$ in \eqref{eq:channel_corr}.
This integral can be computed via the \ac{GL} quadrature \cite{olver10} which takes the form
\vspace{-1ex}
\begin{equation}
\label{eq:Gauss_legendre_general}
\vspace{-1ex}
\int_{\bar{a}}^{\bar{b}} \bar{\psi}(\bar{x}) \, d\bar{x} \approx \frac{\bar{b} - \bar{a}}{2} \sum_{\bar{m}=1}^{\bar{M}} \bar{\omega}_{\bar{m}} \bar{\psi} \left( \frac{\bar{b} - \bar{a}}{2} \bar{\theta}_{\bar{m}} + \frac{\bar{a} + \bar{b}}{2} \right),
\end{equation}
where $\bar{M}$ denotes the number of sample points, $\bar{\omega}_{\bar{m}}$ are the quadrature weights, and $\bar{\theta}_{\bar{m}}$ define the roots of the $\bar{M}$-th Legendre polynomial. 
As seen from \eqref{eq:Gauss_legendre_general}, a larger value of $\bar{M}$ results in a higher approximation accuracy for the integral.

Let $L_x$ and $L_z$ denote the length of the \ac{FCAPA} along the $x$- and $z$-axes, respectively.
Then each $q_{i,m}$-th entry can be calculated as
\vspace{-1ex}
\begin{align}
\label{eq:MF_solution_TX_Gauss_1}
\vspace{-1ex}
q_{i,m} &=  \int_{\mathcal{U}} H_m\big(\mathbf{s}(u,v)\big) H_i^{\ast}\big(\mathbf{s}(u,v)\big)\, {\zeta(u,v)} du \, dv \\
&=  \int_{-\frac{L_z}{2}}^{\frac{L_z}{2}} \int_{-\frac{L_x}{2}}^{\frac{L_x}{2}} H_m(s_x,s_z) H_i^{\ast}(s_x,s_z)\, {\zeta(s_x,s_z)} ds_x \, ds_z \nonumber \\
&\approx  \frac{L_x L_z}{4} \sum_{\bar{m}_z=1}^{\bar{M}} \sum_{\bar{m}_x=1}^{\bar{M}} \omega_{\bar{m}_x} \omega_{\bar{m}_z} H_m\bigg(\frac{L_x \bar{\theta}_{\bar{m}_x}}{2}, \frac{L_z \bar{\theta}_{\bar{m}_z}}{2}\bigg) \nonumber \\ 
&\hspace{8ex}\times H_i^{\ast}\bigg(\frac{L_x \bar{\theta}_{\bar{m}_x}}{2}, \frac{L_z \bar{\theta}_{\bar{m}_z}}{2}\bigg) \zeta\bigg(\frac{L_x \bar{\theta}_{\bar{m}_x}}{2}, \frac{L_z \bar{\theta}_{\bar{m}_z}}{2}\bigg). \nonumber
\vspace{-0.5ex}
\end{align}

Finally, under the assumption that $\frac{\partial \mathcal{F}_k}{\partial g} = 0$, the dominant term $G_d(u,v)$ in \eqref{eq:euler_lagrange_final} can be concisely collected into a matrix $\mathbf{G}_d \in \mathbb{C}^{\bar{M} \times K}$ in an integral-free manner as
\vspace{-1ex}
\begin{align}
    \label{eq:dominant_G_term}
    \vspace{-1ex}
    \mathbf{G}_d &\triangleq 2\,\Re\!\{A_k H_k(u,v) J_k^\star(\cdot;g)\} \\
    &\hspace{-4ex}- {2} \sum_{i=1}^{K} \Big( B_i\,{\Re\!\{ I_{k,i}^*[g] H_k(u,v) J_i^\star(\cdot;g)\}} + C_i\,|J_i^\star(\cdot;g)|^2 \Big) \nonumber \\
    &\hspace{-4ex}= 2 \Re\{ \mathbf{H} \!\odot\! \tilde{\mathbf{J}} \!\odot\! \tilde{\mathbf{a}} \} \!-\! 2 \Big( \Re\{ \mathbf{H} \!\odot\! \big( \tilde{\mathbf{J}} ( \tilde{\mathbf{B}} (\mathbf{W}^*)\trans ) \big) \} \!+\! (\tilde{\mathbf{J}} \!\odot\! \tilde{\mathbf{J}}^*) \mathbf{c} \Big), \nonumber 
\end{align}
where $\odot$ denotes the Hadamard product, $\mathbf{W}^*$ denotes the element-wise conjugate of a matrix $\mathbf{W}$ with the definitions
\vspace{-1ex}
\begin{subequations}
\vspace{-1ex}
    \begin{equation}
        \tilde{\mathbf{a}} \triangleq [A_1, \cdots, A_k, \cdots, A_K] \in \mathbb{C}^{1 \times K},
    \end{equation}
    \begin{equation}
        \tilde{\mathbf{B}} \triangleq \text{diag}([B_1, \cdots, B_i, \cdots, B_K]) \in \mathbb{C}^{K \times K},
    \end{equation}
    \begin{equation}
        \mathbf{c} \triangleq [C_1, \cdots, C_i, \cdots, C_K]\trans \in \mathbb{C}^{K \times 1}.
    \end{equation}
    \label{eq:Grad_aux_updates}
\end{subequations}

\vspace{-1.5ex}
\begin{algorithm}[H]
\caption{FCAPA Optimization for WSR Maximization}
\label{alg:proposed_decoder}
\setlength{\baselineskip}{11pt}
\textbf{Input:} Iterations $i_{\mathrm{S}}$, transmit power $P_\mathrm{T}$, aperture size $A_\mathrm{T}$, surface shape $g(u,v)$ and morphability factor $\xi$. \\
\textbf{Output:} Optimal source currents $J_k^\star\big(\mathbf{s}(u,v);g\big), \forall k$ and optimal surface shape $g^\star(u,v)$. 
\vspace{-1.5ex} 
\begin{algorithmic}[1]  
\STATEx \hspace{-3.5ex}\hrulefill
\STATEx \hspace{-3.5ex}\textbf{Initialization}
\STATEx \hspace{-3.5ex} - Set source currents $J_k^\star\big(\mathbf{s}(u,v);g\big) = H_k^{\ast}\big(\mathbf{s}(u,v)\big)$.
\STATEx \hspace{-3.5ex} - Set auxiliary variables $\mu_k^{(0)} = \mu_k(\mathbf{Q})$ and $\lambda_k^{(0)} = \lambda_k(\mathbf{Q})$.
\STATEx \hspace{-3.5ex} - Choose surface shape $g(u,v)$.
\vspace{-1ex}
\STATEx \hspace{-3.5ex}\hrulefill
\STATEx \hspace{-3.5ex}\textbf{for} $i=1$ to $i_{\mathrm{S}}$ \textbf{do}:
\STATE Update $\mathbf{A}$, $\mathbf{B}$, $\tilde{\mathbf{a}}$, $\tilde{\mathbf{B}}$ and $\mathbf{c}$ via \eqref{eq:A_def}, \eqref{eq:B_def} and \eqref{eq:Grad_aux_updates}.
\STATE Update $\mathbf{W}$ via \eqref{eq:final_w}.
\STATE Update $\mu_k, \forall k$ and $\lambda_k, \forall k$ via \eqref{eq:muk_sol_W} and \eqref{eq:lambdak_sol_W}.
\STATE Calculate $G(u,v)$ via \eqref{eq:euler_lagrange_final} and \eqref{eq:dominant_G_term}. 
\STATE Update $g(u,v)$ and its derivatives via \eqref{eq:Ful_gradeint_update}.
\STATE Update $\mathbf{Q}$ via \eqref{eq:MF_solution_TX_Gauss_1}.
\STATEx \hspace{-3.5ex}\textbf{end for}
\STATE Calculate optimal $J_k^\star\big(\mathbf{s}(u,v);g\big), \forall k$ via \eqref{eq:Jk_final_uv}.
\STATE Normalize $J_k^\star\big(\mathbf{s}(u,v);g\big), \forall k$ via \eqref{eq:deformed_obj_uncon_sol}.

\end{algorithmic}
\end{algorithm}
\vspace{-2ex}

Note that as a result of the matrix operations, $\mathbf{G}_d$ is a $\bar{M} \times K$ complex matrix since both $\mathbf{H}$ and $\tilde{\mathbf{J}}$ are also defined as a $\bar{M} \times K$ complex matrices, where $\mathbf{H}$ can be obtained via the previous \ac{GL} quadrature method and $\tilde{\mathbf{J}}$ can be calculated as
\vspace{-1ex}
\begin{equation}
\vspace{-1ex}
    \tilde{\mathbf{J}} = \mathbf{H}^* (\mathbf{A} - \mathbf{B} \mathbf{W}\trans) \in \mathbb{C}^{\bar{M} \times K}.
\end{equation}

Finally, representing the $k$-th column of $\mathbf{G}_d$ as $\mathbf{G}_d^{(k)}$ lets us extract $\mathbf{g}_d \triangleq \sum_{k=1}^K \mathbf{G}_d^{(k)} \in \mathbb{C}^{\bar{M} \times 1}$ where each $(u,v)$-th element of $\mathbf{g}_d$ is $G_d(u,v)$.

The complete procedure is summarized in Algorithm \ref{alg:proposed_decoder}.

\section{Performance Analysis}
\label{sec:perf_analy}

Let $\xi$ denote the maximum admissible ``morphability'' of the \ac{FCAPA} such that
\vspace{-1ex}
\begin{equation}
\vspace{-1ex}
g(u,v) - \frac{\xi}{2} \leq g(u,v) \leq g(u,v) + \frac{\xi}{2},
\end{equation}
leading to $y_\text{max} \triangleq g(u,v) + \frac{\xi}{2}$ and $y_\text{min} \triangleq g(u,v) - \frac{\xi}{2}$.

To evaluate the performance gains of the proposed \ac{FCAPA} system, we assume a paraboloid surface given as
\vspace{-1ex}
\begin{equation}
    \label{eq:chosen_shape}
    \vspace{-1ex}
    g(u,v) = u^2 + v^2.
\end{equation}

\begin{remark}
    The choice of surface shape is completely arbitrary in this section, but it can be chosen for a specific purpose \cite{Antenna_theory}. 
    For example, paraboloid surfaces \cite{Raj2012} are typically used for focusing signals in communications and radar systems.
\end{remark}

For simplicity, we also assume that the explicit dependence of $H_k(u,v)$ on $g$ is negligible due to the far-field assumption; $i.e., \frac{\partial \mathcal{F}_k}{\partial g} \approx 0$.
The inclusion of this term requires derivatives of the Green's function, which, while possible, are computationally intensive and intractable.

Following \eqref{eq:chosen_shape}, the transmit \ac{FCAPA}'s deployment region within the $x$-$y$-$z$ plane can be expressed as
\vspace{-1ex}
\begin{equation}
    \label{eq:surface_def_numerical}
    \vspace{-1ex}
    \mathcal{S} = \bigg\{ \big[s_x, s_x^2 + s_z^2, s_z\big]\trans \Big| |s_x| \leq \frac{L_x}{2}, |s_z| \leq \frac{L_z}{2}  \bigg\},
\end{equation}
where $L_x = L_z = \sqrt{A_\mathrm{T}}$, with $A_\mathrm{T}$ denoting the aperture size.

\begin{figure}[H]
    \centering
    \includegraphics[width=\columnwidth]{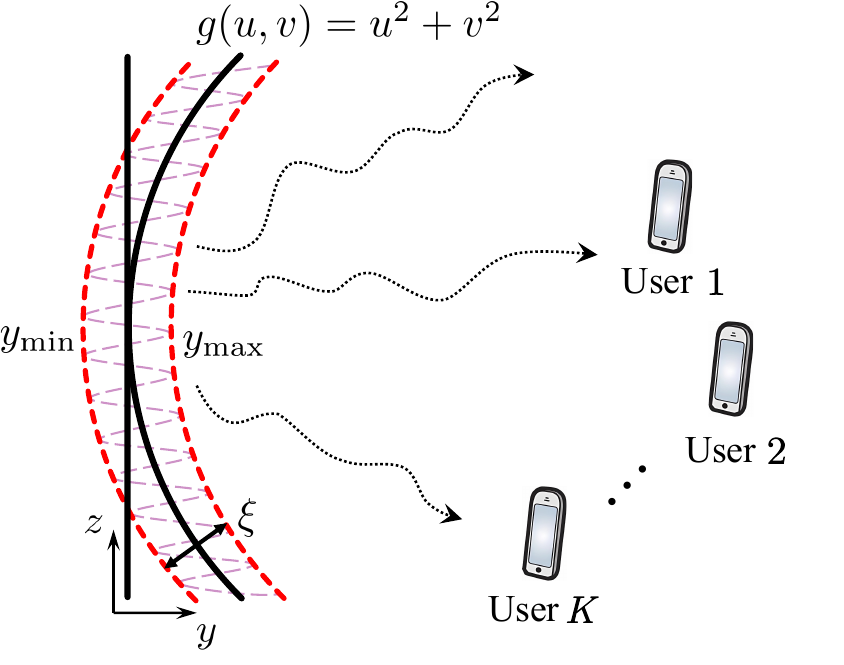}
    \caption{Cross-section of the simulated \ac{FCAPA} model with $g(u,v) = u^2 + v^2$ and the morphing range $\xi$.
    Each antenna element is initially arranged on the black curve and then adjusted according to the optimization algorithm with the limits of morphing defined by the red dashed curves.
    An example \ac{FCAPA} shape after optimization is shown with the transparent purple dashed line between the red curves.}
    \label{fig:FCAPA_simulated}
\end{figure}

A visual overview of the shape is presented in Fig. \ref{fig:FCAPA_simulated}.
Unless otherwise specified, the default aperture size is set to $A_\mathrm{T} = 0.25$ m$^2$.

\vspace{-2ex}

We also assume that there are $K = 8$ communications users randomly and uniformly distributed within the region
\vspace{-1ex}
\begin{equation}
    \label{eq:surface_def_numerical_users}
    \!\!\!\!\!\mathcal{K} \!=\! \bigg\{ \!\big[r_x,\!r_y,\!r_z\big]\!\trans \Big| |r_x|\! \leq\! R_x, |r_z|\! \leq\! R_z,\! R_y^\text{min}\! \leq \!r_y\! \leq \!R_y^\text{max}  \!\bigg\}\!,\!
\vspace{-1ex}
\end{equation}
where $R_x = R_z = 5$ m, $R_y^\text{min} = 15$ m and $R_y^\text{max} = 30$ m.

In addition, each user is assumed to have a polarization direction specified by $\hat{\mathbf{u}}_x = \hat{\mathbf{u}}_z = [0, 0, 1]\trans, \forall k$.
The rest of the simulation parameters are summarized in Table \ref{table:sim_params}.

Since the user weight is considered to be identical for all the users, the numerical results are hereafter denoted in terms of the \ac{ARPU} for clarity.

The proposed \ac{BCD}-\ac{CoV}-\ac{EL} method for \ac{FCAPA} is also compared against the following benchmarking schemes.

\noindent \textbf{BCD-CoV for CAPA}: For this benchmark, we adopt the \ac{BCD}-\ac{CoV} method proposed in \cite{WangTWC2025} to optimize a typical \ac{CAPA} with low complexity.

\noindent \textbf{Fourier method for CAPA}: Similarly, this benchmark is a higher complexity alternative to the \ac{BCD}-\ac{CoV} approach where the continuous source current patterns are approximated using a finite Fourier series \cite{WangTWC2025}.
Since this method has an intrinsically higher complexity which increases with aperture size and/or signal frequency, we do not derive its equivalent for the \ac{FCAPA} model.

\noindent \textbf{Flexible MIMO}: This benchmark is the flexible variant of a spatially discrete antenna array with an identical shape $g(u,v)$ (and a correspondingly identical morphing range $\xi$) to the \ac{FCAPA} model which resembles the closest comparison to a regular \ac{FIM}.
For this model, we assume that the continuous surface $\mathcal{S}$ is occupied with discrete antenna elements where each antenna element occupies an area of $A_d = \frac{\lambda^2}{4\pi}$ m$^2$ with an antenna spacing of $d = \frac{\lambda}{2}$.

\begin{table}[H]
\centering
\caption{System Parameters}
\vspace{-2ex}
\label{table:sim_params}
\begin{tabular}{|c|c|c|}
\hline
\textbf{Parameter} & \textbf{Symbol} & \textbf{Value} \\
\hline
Signal frequency & $f$ & $2.4$ GHz \\
\hline
Intrinsic impedance & $\eta$ & $120\pi$ $\Omega$ \\
\hline
Transmit power factor\footnotemark & $P_\mathrm{T}$ & $100$ mA$^2$ \\
\hline
Noise power factor & $\sigma^2_k$ & $5.6 \times 10^{-3}$ V$^2$/m$^2$ \\
\hline
User weight & $\alpha_k$ & $1/K, \forall k$ \\
\hline
GL quadrature samples & $\bar{M}$ & $20$ \\
\hline
Monte Carlo realizations & - & $200$ \\
\hline
\end{tabular}
\vspace{-2ex}
\end{table}

Then, the location of the $(n_x, n_z)$-th antenna can be expressed as
\vspace{-2ex}
\begin{equation}
       \bar{\mathbf{s}}_{n_x, n_z} = \bigg[ (n_x - 1) d - \frac{L_x}{2}, \bar{g}, (n_z - 1) d - \frac{L_z}{2} \bigg]\trans,
\vspace{-1ex}
\end{equation}
where $\bar{g} \triangleq \big( (n_x - 1) d - \tfrac{L_x}{2} \big)^2 + \big( (n_z - 1) d - \tfrac{L_z}{2} \big)^2$ when \eqref{eq:chosen_shape} is used.

This discretization yields a total of $N_d = \lceil \tfrac{L_x}{d} \rceil \times \lceil \tfrac{L_z}{d} \rceil$ discrete antennas.
Next, let $\mathcal{S}_{n_x, n_z}$ denote the total surface of the $(n_x, n_z)$-th antenna, where $|\mathcal{S}_{n_x, n_z}| = A_d$.
Then, the channel between each $(n_x, n_z)$-th antenna and user $k$ can be calculated as
\vspace{-1ex}
\begin{align}
  h_{k, n_x, n_z} &= \frac{1}{\sqrt{A_d}} \int_{\mathcal{S}_{n_x, n_z}} \!\!\!\!\!\!\!\!\!\!\!\!\!H_k\big(\mathbf{s}(u,v)\big)\,\zeta(u,v)\; du\,dv \nonumber \\
  &\approx \sqrt{A_d} H_k\big(\bar{\mathbf{s}}_{n_x, n_z}\big)\,\zeta(n_x,n_y).
\vspace{-2ex}
\end{align}

Leveraging the discrete channels above, one can now optimize this traditional \ac{MIMO} beamforming optimization problem with typical \ac{SotA} methods such as a low-complexity zero forcing or a more advanced fractional programming approach.

\noindent \textbf{Conventional MIMO}: This benchmarking scheme is similar to that above except that we now consider no morphabilty and hence, $\bar{g} =0$ and $h_{k, n_x, n_z}$ no longer contains $\zeta(n_x,n_y)$.

\vspace{-2ex}
\subsection{Numerical Results}

\begin{figure}[H]
\includegraphics[width=\columnwidth]{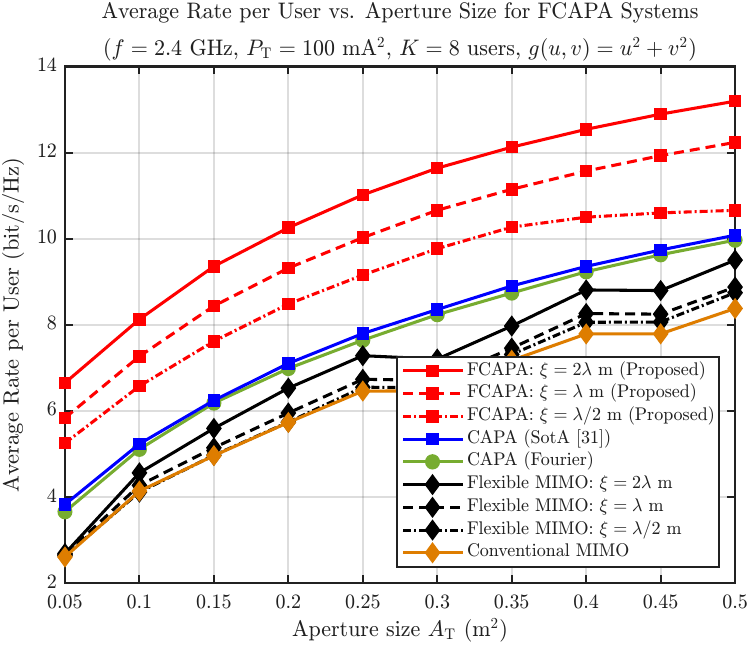}
\caption{\ac{ARPU} performance of the proposed \ac{FCAPA} system with a varying aperture size compared to the \ac{SotA}.}
\label{fig:aperture_vary}
\end{figure}

Let us first analyze how the \ac{ARPU} varies with an increasing aperture size.
As seen from Fig. \ref{fig:aperture_vary}, there is a significant increase in the \ac{ARPU} at all aperture sizes compared to the \ac{SotA} when using \ac{FCAPA}.
\vspace{-1ex}

\footnotetext{Since antenna efficiency is neglected and the transmit power is directly proportional to the source current density, the power-related terms are described in terms of \emph{power factors}, with the unit A$^2$ (see \cite[page 526, eq.~(15.2.3)]{orfanidis2008electromagnetic}), conforming to physical system and \ac{SINR} definitions \cite{WangTWC2025,wang2025beamformingdesigncontinuousaperture,WangTCOM2025,MarzettaJSAIT2025}.}

\begin{figure}[H]
\includegraphics[width=\columnwidth]{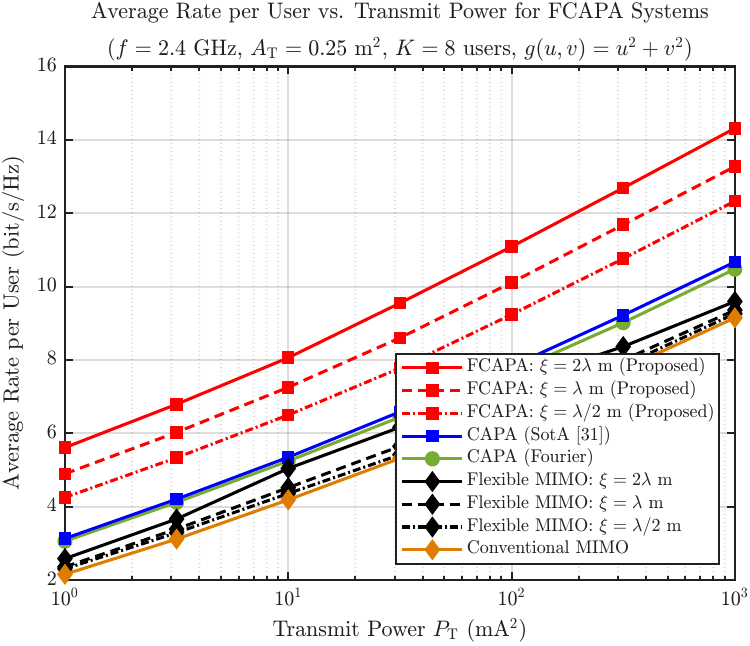}
\caption{\ac{ARPU} performance of the proposed \ac{FCAPA} system with a varying transmit power compared to the \ac{SotA}.}
\label{fig:transmit_power_vary}
\vspace{2ex}
\includegraphics[width=\columnwidth]{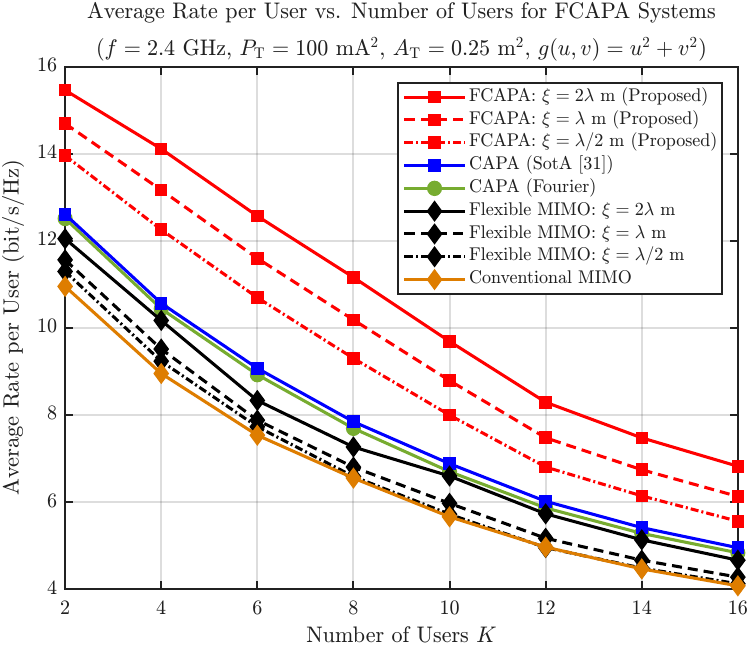}
\caption{\ac{ARPU} performance of the proposed \ac{FCAPA} system with a varying number of users compared to the \ac{SotA}.}
\label{fig:users_vary}
\vspace{-2ex}
\end{figure}

As expected, the \ac{ARPU} increases with a larger morphability $\xi$ since there are naturally more \acp{DoF} to be exploited.
Additionally, while the compared flexible \ac{MIMO} scheme outperforms the conventional \ac{MIMO} setup, the performance does not quite reach the threshold seen when using typical \ac{CAPA} systems.

Next, Fig. \ref{fig:transmit_power_vary} shows the variation in the \ac{ARPU} when the transmit power is varied.
As illustrated, there is a similar trend with the proposed \ac{FCAPA} system outperforming all the aforementioned \acp{SotA}, with an increase in \ac{ARPU} seen when the transmit power is increased.
In addition, the variation with a changing $\xi$ also remains consistent as with Fig. \ref{fig:aperture_vary}.

\begin{figure}[H]
\includegraphics[width=\columnwidth]{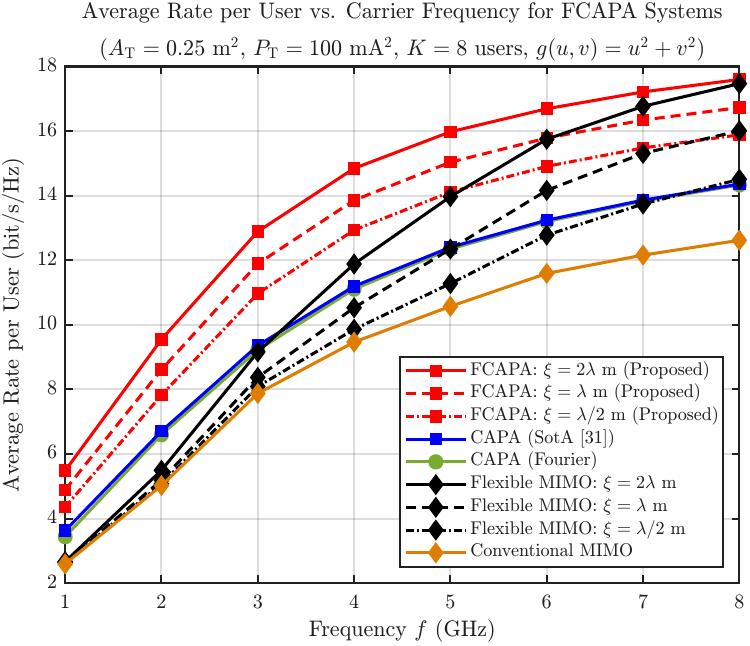}
\vspace{-3ex}
\caption{\ac{ARPU} performance of the proposed \ac{FCAPA} system with a varying carrier frequency compared to the \ac{SotA}.}
\label{fig:freq_vary}
\vspace{1ex}
\includegraphics[width=\columnwidth]{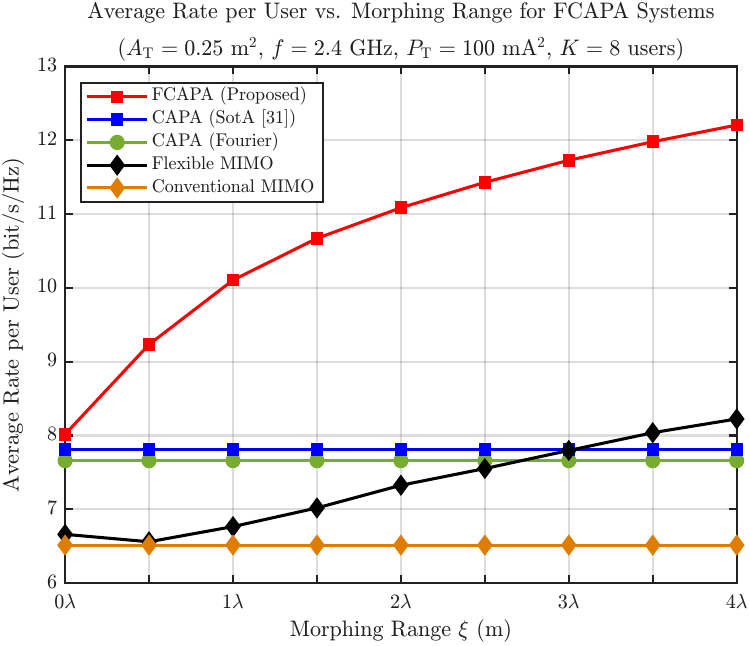}
\vspace{-3ex}
\caption{\ac{ARPU} performance of the proposed \ac{FCAPA} system with a varying morphing range $\xi$.}
\vspace{-3ex}
\label{fig:xi_vary}
\end{figure}

Next, Figs. \ref{fig:users_vary} and \ref{fig:freq_vary} portray the variation of the \ac{ARPU} with an increasing number of users and carrier frequencies, respectively.
As expected, the trend continues with the proposed \ac{FCAPA} system outperforming all the \acp{SotA} with a common decrease in the \ac{ARPU} seen with an increasing number of users and an increase in the \ac{ARPU} seen when the carrier frequency is increased.
It is also noteworthy that the flexible \ac{MIMO} variant has a higher \ac{ARPU} reaching the performance of the proposed \ac{FCAPA} at higher carrier frequencies.

Finally, Fig. \ref{fig:xi_vary} portrays the variation in the \ac{ARPU} when the morphing range $\xi$ is increased.
As expected, when there is no morphing for a given shape; $i.e., \xi = 0$, the performance of an \ac{FCAPA} system is identical to that of a typical \ac{CAPA}.
Another interesting trade-off that can be seen from the figure is the crossover point at $\xi = 3\lambda$ when both the typical \ac{CAPA} and flexible \ac{MIMO} variations have the same \ac{ARPU}.
\newpage

\begin{figure}[H]
\includegraphics[width=\columnwidth]{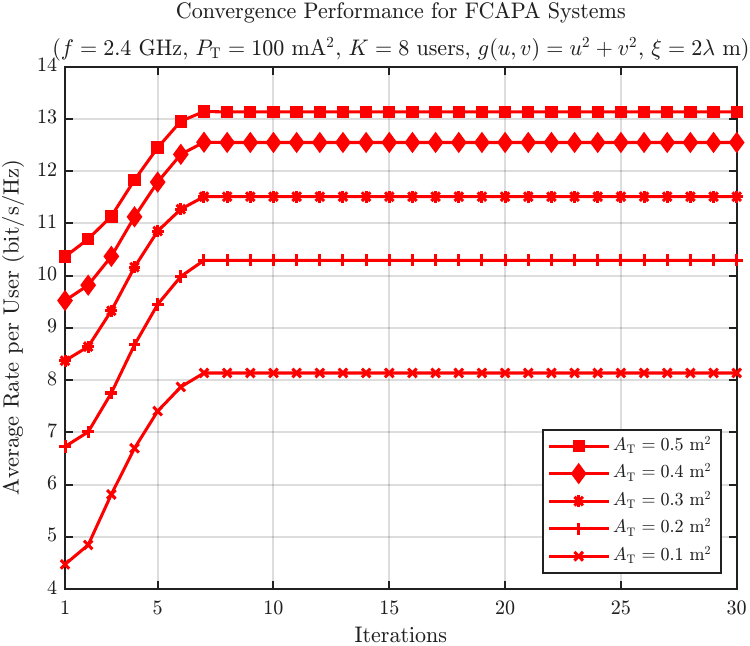}
\caption{Convergence behavior for the proposed \ac{FCAPA} optimization in Algorithm \ref{alg:proposed_decoder}.}
\label{fig:convergence}
\end{figure}

\subsection{Convergence and Complexity}

We now present a convergence plot for the proposed Algorithm \ref{alg:proposed_decoder} in Fig. \ref{fig:convergence}.
As seen from the figure, the proposed technique converges in less than $10$ iterations for all the aperture sizes.
Although one cannot guarantee convergence to a global maximum due to the gradient-based approach leveraged, the algorithm converges to local maxima in relatively few steps, demonstrating the effectiveness of the proposed approach.

The computational complexity of the proposed method is dominated by the matrix inversion required to execute equation \eqref{eq:final_w}, which amounts to $\mathcal{O}(K^3)$.
Asymptotically, this complexity is identical to that of the regular \ac{CoV}-based approach used in \cite{WangTWC2025} since the computation of the gradients and their derivatives have a lower complexity than the matrix inversion.

\section{Conclusion}
\label{sec:conclusion}

In conclusion, we introduced a novel \ac{EM} architecture, termed \textit{flexible \ac{CAPA} (\ac{FCAPA})}, which integrates intrinsic surface flexibility into conventional \ac{CAPA} systems to fully exploit the available \ac{DoF} in \ac{MIMO} systems.
Through the formulation and solution of a downlink multi-user beamforming optimization problem aimed at maximizing the \ac{WSR}, we demonstrated that the proposed \ac{FCAPA} structure significantly outperforms traditional \ac{CAPA} configurations, with performance gains that grow proportionally to the degree of surface morphability.

\bibliographystyle{IEEEtran}
\bibliography{references}

\end{document}